\documentclass[numbook]{svjour2}                    
\smartqed  
\usepackage{graphicx,amsmath,amssymb,bm,subfigure}
\usepackage{amsmath}
\usepackage{slashed}
\usepackage{rotating}
\usepackage{multirow}
\usepackage{latexsym}
\usepackage{textcomp}
\usepackage{verbatim}
\usepackage{color}
\usepackage{pict2e}
\usepackage{bm}
\usepackage{everysel}
\usepackage{keyval}
\usepackage{ragged2e}
\usepackage{dsfont}
\usepackage{amssymb}
\usepackage{enumitem}
\usepackage{pstricks}
\usepackage{mathtools}
\usepackage{hyperref}

\def\s12{spin-${\frac{1}{2}}$}

\def \overleftright#1{#1 {\kern-8pt\hbox{{\raise10pt\hbox
{$\scriptstyle{\leftrightarrow}$}}}}}

\begin{document}

\title{Disorder Operators and their Descendants}

\author{Eduardo Fradkin}

\institute{Department of Physics and Institute for Condensed Matter Theory, \at
              University of Illinois, 1110 West Green Street, Urbana, Illinois 61801-3080, USA\\
              \email{efradkin@illinois.edu}   }
                      
\date{Received: date / Accepted: date}

\maketitle

\begin{abstract}
I review the concept of a {\em disorder operator}, introduced originally by Kadanoff
 in the context of the two-dimensional Ising model. Disorder operators  acquire an expectation value in the disordered phase of the classical spin system. This concept has had applications and implications to many areas of physics ranging from quantum spin chains to gauge theories to topological phases of matter. In this paper I  describe the role that  disorder operators play in our understanding of ordered, disordered and topological phases of matter. The role of disorder operators, and their generalizations, and their connection with dualities in different systems, as well as with majorana fermions and parafermions,  is discussed in detail. Their role  in recent fermion-boson and boson-boson dualities is briefly discussed.
\end{abstract}

\tableofcontents
\section{Introduction}
\label{intro}

The phases of matter of many physical systems can be labeled (or classified) by the symmetries of their local order parameters. In this picture the physical system has a global symmetry and the order parameter is a local observable that transforms under some irreducible representation of this symmetry \cite{Landau-Lifshitz-StatPhys}. In the ordered phase the order parameter has a non-vanishing expectation and the global symmetry is broken spontaneously. 

However, there are systems without any broken symmetries which can be regarded as condensates of seemingly non-local operators. 
This concept originates in the pioneering  work of Leo Kadanoff  (with his then student Horacio Ceva) of 1971 in which the concept of disorder operator 
was first introduced: an operator that has a vanishing expectation value in the broken symmetry phase of the 2D Ising model 
but   has a non-vanishing expectation value in the disordered phase. 
Such disorder operators have since been found in diverse systems including gauge theories and topological phases of quantum antiferromagnets. 
In this paper I  review the Kadanoff-Ceva construction and several of its notable extensions to diverse systems. 
Here I discuss the role of disorder operators to our understanding of phases of matter and topological states. I also discuss the concept of particle-vortex duality and its many recent extensions, including fermion-boson dualities.

The paper is organized as follows. In Section \ref{sec:2D-ising}, I present the Kadanoff-Ceva construction of disorder operators in the classical 2D Ising model and its connection with  Onsager fermions, and with Kramers-Wannier duality. In Section \ref{sec:1D-TFIM}, disorder operators (kinks)are constructed in the 1D quantum Ising model (the Ising model in a transverse field). Here their relation with Majorana fermions (and the Jordan-Wigner transformation) and with the quantum version of Ising duality (including a discussion of Majorana zero modes) is discussed. The extension of these constructions to the 2D classical and 1D quantum $\mathbb{Z}_N$ clock models, and the associated parafermions and particle-vortex duality in both models, is discussed in Section \ref{sec:clock}. Here I present a brief discussion of parafermion zero modes and their role as platforms for topological quantum computation. The role of disorder operators in  topological (Haldane) spin chains in discussed in Section \ref{sec:spin-chains}. In Section \ref{sec:IGT}, I discuss Ising gauge theory in 2+1 dimensions (both as a 3D Euclidean lattice and as a 2D Hamiltonian formulation). The role of disorder operators as $\mathbb{Z}_2$ monopole operators,  in  confining and  deconfined (topological) phases, and the effective topological field theory (Chern-Simons) description of the latter phase is discussed in this Section. Particle-vortex dualities in 3D classical and 2+1-dimensional quantum systems, including the 3D classical $XY$ model, loop models of the superconductor-insulator transition, and quantum Hall fluids, are discussed in Section \ref{sec:generalized}. A loop model with an extended complex duality to an $SL(2,\mathbb{Z})$ modular symmetry is discussed. The connection with recent boson-boson and fermion-boson dualities in various dimensions and their role in the physics of topological insulators is also presented here. 

The introduction of the concept of a disorder operator in Kadanoff's pioneering 1971 paper \cite{Kadanoff-1971}  had long-lasting implications for our understanding of phases of matter. This concept  plays a key role in wide areas of physics, ranging from classical and quantum spin systems, topological phases of matter and  gauge theory. The aim of this paper is to honor the work that  Leo  did and his role in shaping how we think about these fascinating problems. I am forever indebted to him  for his deep insights, generosity and friendship. Leo had great influence on the way I think about physics since the early stages of my career to the present time. I will miss him. 

\section{Disorder operators in the classical 2D Ising model}
\label{sec:2D-ising}

The prototype of this picture is the two-dimensional Ising model of a ferromagnet with uniaxial anisotropy. In the ferromagnetic Ising model at each site $\bm{r}$ of the square lattice there is a magnetic moment that can take two possible values, $\sigma(\bm r)=\pm 1$, corresponding to the up and down spin states. The energy of a configuration of spins $[\sigma]$ is 
\begin{equation}
E[\sigma]=- J \sum_{\langle {\bm r}, {\bm r}'\rangle} \sigma(\bm r) \sigma({\bm r}')
\label{eq:Ising}
\end{equation}
where the sum runs over nearest-neighboring sites of the lattice, denoted  by $\langle {\bm r}, {\bm r}'\rangle$. In dimensions $d>1$, the Ising model has a stable ordered phase at temperatures below critical temperature
and a disordered phase for $T>T_c$. In the ordered phase that order parameter has a uniform non-vanishing  expectation value $m=\langle \sigma(\bm r)\rangle \neq 0$, which vanishes in the disordered phase, $m=\langle \sigma(\bm r) \rangle=0$.

\subsection{The Kadanoff-Ceva disorder operator}
\label{sec:kadanoff-ceva}

\begin{figure}[hbt]
\begin{center}
\subfigure[]{
  \includegraphics[width=0.48\textwidth]{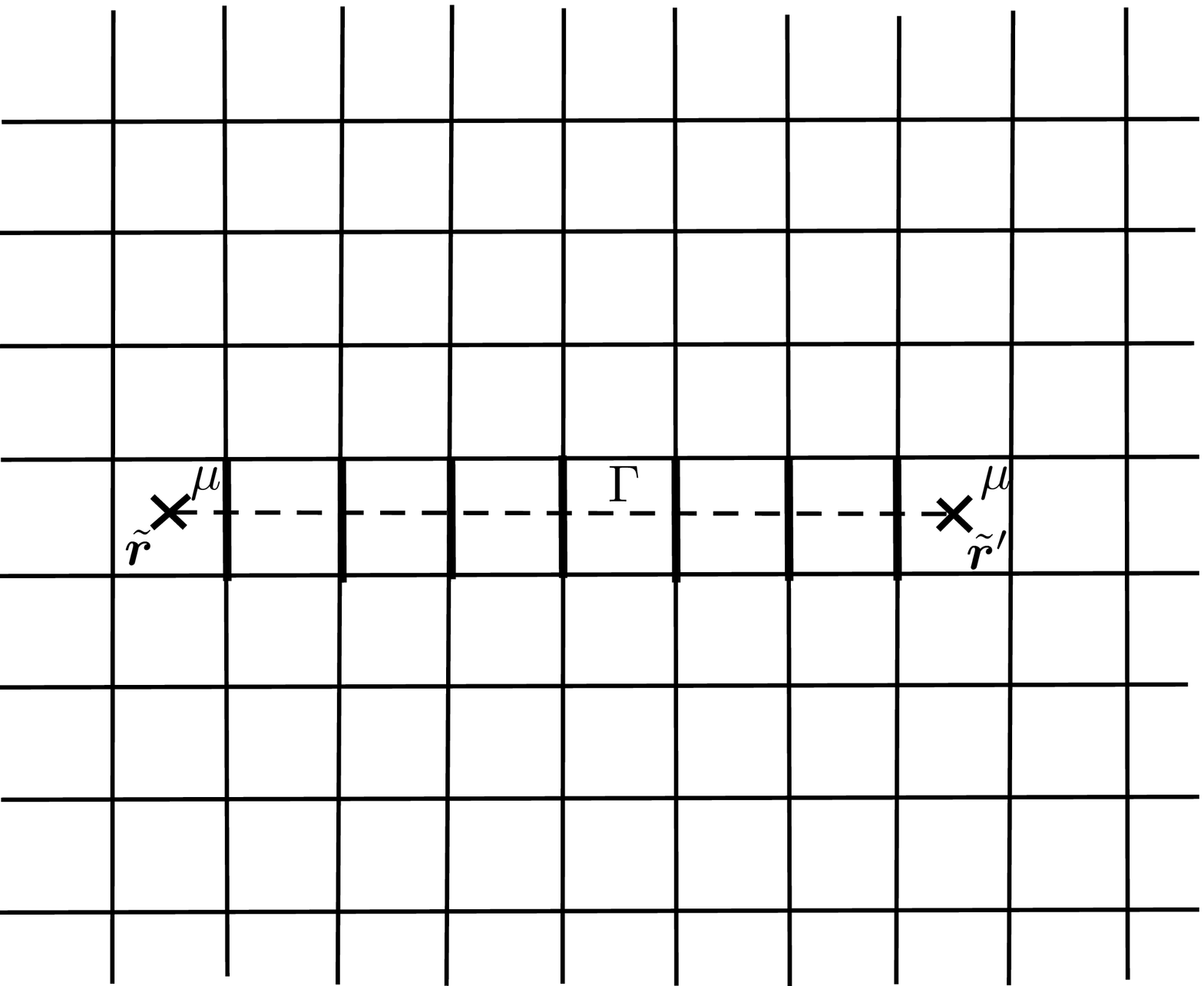}}
  \subfigure[]{ 
  \includegraphics[width=0.48\textwidth]{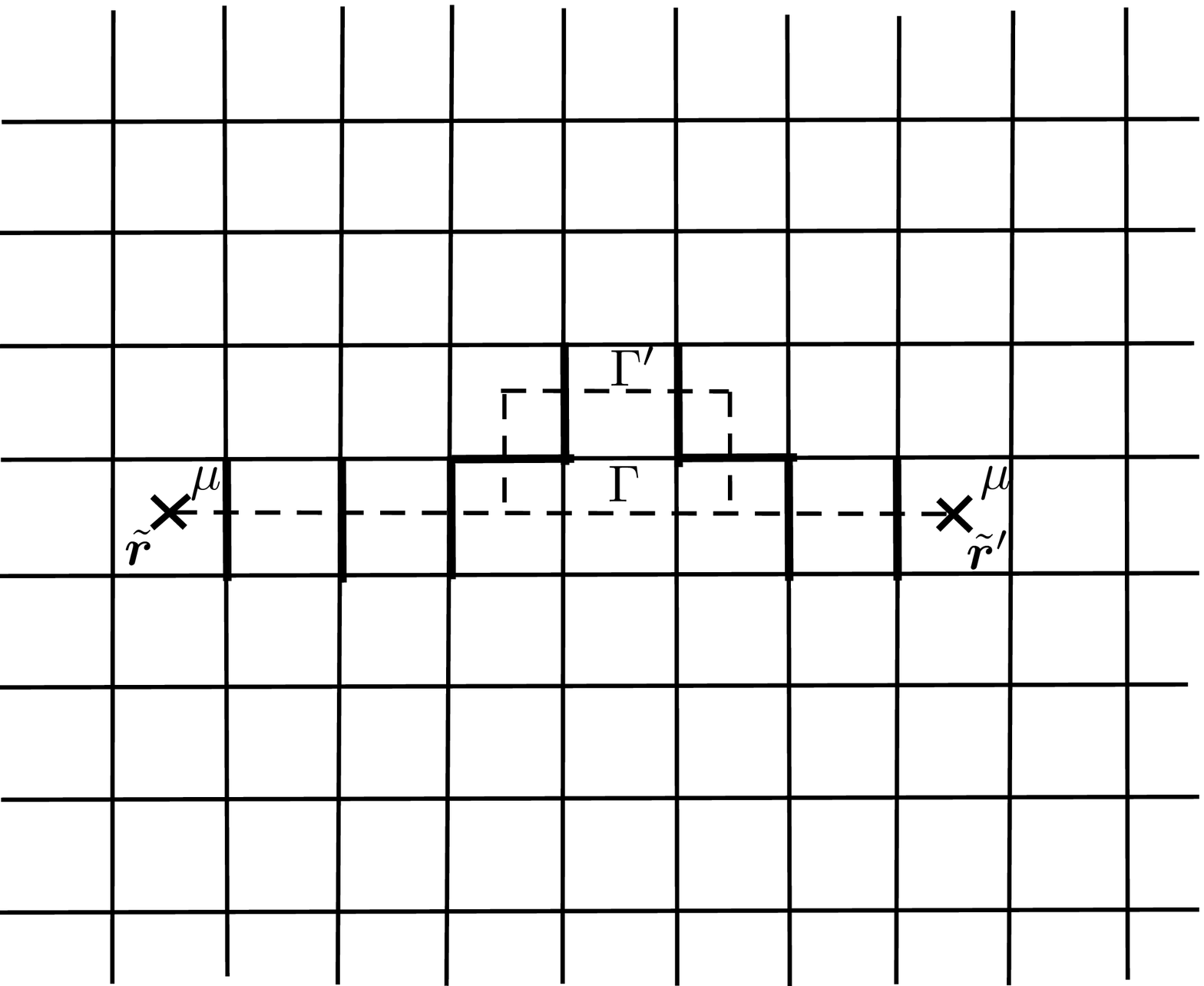}}
\caption{Disorder Operators: (a) The broken path $\Gamma$ is a path of the dual lattice, spanning the dual sites ${\tilde {\bm r}}$ and ${\tilde {\bm r}}'$, along which the bonds of the direct lattice have antiferromagnetic sign. This configuration of bonds represents two disorder operators, $\mu({\tilde {\bm r}})$ and $\mu({\tilde {\bm r}}')$ defined on the  sites ${\tilde {\bm r}}$ and ${\tilde {\bm r}}'$ of the dual lattice. (b) Disorder operators are the same if the path $\Gamma$ is distorted to another path $\Gamma'$ spanning the same pair of dual sites.}
\end{center}
\label{fig:disorder-operators}       
\end{figure}

In 1971 Kadanoff and Ceva\cite{Kadanoff-1971} introduced the concept of {\em disorder variables} (or {\em disorder operators}) in the context two-dimensional Ising model, i.e. operators that acquire an expectation value in the disordered phase of a classical spin system. Let ${\tilde {\bm r}}$ and ${\tilde {\bm r}}'$ denote two sites of the dual lattice (see Fig. \ref{fig:disorder-operators}a and b). Kadanoff and Ceva proposed to consider a modified 2D ferromagnetic Ising model with a defect that creates a fractional domain wall along a path $\Gamma$ of the dual lattice. To introduce this defect is the same as to changing the coupling constant $J$ to have an antiferromagnetic (negative) sign along a seam of bonds pierced by the path $\Gamma$ (denoted by the set of bold links in Fig. \ref{fig:disorder-operators}a). The correlation function of the two disorder operators is defined to be
\begin{equation}
\langle \mu({\tilde {\bm r}}) \mu({\tilde {\bm r}}')\rangle=\frac{Z[\Gamma]}{Z}\equiv \exp(-\Delta F[\Gamma]/T)
\label{eq:mumu}
\end{equation}
Here $Z$ is the partition function of the 2D ferromagnetic Ising model, $Z[\Gamma]$ is the partition function with the fractional domain wall, and $\Delta F[\Gamma]$ is the excess free energy caused by the defect. 

Deep in the ferromagnetic phase, in the presence of this defect  there is an  excess free energy originating essentially from the spin degrees of freedom within a correlation length of the fractional domain wall. As a result, the excess free energy grows linearly with the separation between the disorder operators, $\Delta F[\Gamma]=\kappa |{\tilde {\bm r}}-{\tilde {\bm r}'}|$. In the thermodynamic limit, and for large separations, $\kappa$ is the same as the  line tension of a 2D domain wall \cite{Fisher-1967}. Hence, in the ordered phase the correlation function of the disorder variables vanishes exponentially fast at long distances. On the other hand, in the disordered phase, where the symmetry is unbroken, the effects of the defect can only be appreciable at a distance of the order of the correlation length of the {\it endpoints} of the path $\Gamma$. Hence, in the disordered phase the correlator of disorder operators approaches a {\em finite} limit at asymptotically large separations, $a \ll |{\tilde {\bm r}}-{\tilde {\bm r}}'|\ll L$ (where $L$ is the linear size of the system):  in the disorder phase the disorder operator has an expectation value. In this sense, the disordered phase can be regarded as condensate of defects. To summarize, the correlator for large separations has the asymptotic behavior
\begin{equation}
\langle \mu({\tilde {\bm r}}) \mu({\tilde {\bm r}}')\rangle=
\begin{cases}
\textrm{const.}\times \dfrac{e^{-\kappa |{\tilde {\bm r}}-{\tilde {\bm r}}'|}}{|{\bm r}-{\bm r}'|^{1/2}},\quad T<T_c \\
\dfrac{\textrm{const.}}{ |{\bm r}-{\bm r}'|^{1/4}}, \quad T=T_c\\
|\langle \mu \rangle|^2+ O(e^{-\kappa' |{\tilde {\bm r}}-{\tilde {\bm r}}'|}), \quad T>T_c
\end{cases}
\label{eq:disorder-correlator}
\end{equation}
where $|\langle \mu \rangle |\propto (T-T_c)^{1/8}$, and $T_c$ is the Onsager critical temperature, $T_c=2J/\ln (\sqrt{2}+1)$.

The configuration of disorder operators is more systematically defined using the language of gauge theory \cite{Wilson-1974,Kogut-1979}  by introducing a set of Ising variables $[\tau({\bm r},{\bm r}')=\pm 1]$ on the links of the lattice, i.e. a set of background Ising gauge fields \cite{Fradkin-1978b}. In this language, the Ising spins are regarded as ``matter fields'' and the variables $[\tau({\bm r},{\bm r}']$ are (background) Ising gauge fields. The partition function $Z[\Gamma]$ for a system with the defect $\Gamma$ now is
\begin{equation}
Z[\Gamma]=\sum_{[\sigma]} \exp\Big[\frac{J}{T} \sum_{\langle {\bm r}, {\bm r}'\rangle} \sigma(\bm r) \tau({\bm r},{\bm r}') \sigma({\bm r}')\Big]
\label{eq:Z_gamma}
\end{equation}
This partition function reproduces that of Eq.\eqref{eq:mumu} for the configuration of Ising gauge fields $\tau({\bm r},{\bm r}')=-1$ for the bonds pierced by the path $\Gamma$ and $\tau({\bm r},{\bm r}')=+1$ for all other bonds. It is then simple to see that the partition function of Eq.\eqref{eq:Z_gamma} is unchanged under the local gauge transformation, $\sigma(\bm r) \mapsto -\sigma(\bm r)$ and $\tau({\bm r}, {\bm r}') \mapsto - \tau({\bm r}, {\bm r}')$ on all bonds $({\bm r}, {\bm r}')$ that share the same site $\bm r$. Furthermore, it is also easy to see that the Wilson loop operator 
\begin{equation}
W_\gamma=\prod_{({\bm r},{\bm r}') \in \gamma} \tau({\bm r}, {\bm r}')
\label{eq:Wilson-loop}
\end{equation}
where $\gamma$ is a closed path of links of the lattice, is gauge-invariant, i.e. unchanged under gauge transformations. The Wilson loop operator takes the value $W_\gamma=-1$ on {\em any} closed loop $\gamma$ that contains the location of just one (but not both) of the disorder operators in its interior, and the value $W_\gamma=+1$ for all other loops. In particular, $W_\gamma=-1$ for the elementary loop (the {\em plaquette}) that contains one of the the disorder operators, since in these plaquettes there is an odd number (one) of links on which the variable $\tau=-1$.  

Upon inspection of the energies of the spin configurations of plaquettes associated with a disorder operator, one can see that the ground state energy of these plaquettes is always larger than $-4J$ since there is always one unsatisfied bond. For this reason such plaquettes are said to be {\em frustrated} \cite{Toulouse-1977}. The concept of geometric frustration played an important role in the physics of spin glasses and in quantum antiferromagnets is at the root of the concept of a quantum spin liquids (for a recent review on  frustration and quantum spin liquids see Ref. \cite{Savary-2016}).

From these observations it follows that the correlator of Eq.\eqref{eq:mumu} is actually {\em path-independent} and depends only on the location of the disorder operators. For instance, the correlators defined by the paths $\Gamma$ and $\Gamma'$ in Fig. \ref{fig:disorder-operators} are exactly equal to each other. This also means that, in spite of their apparently non-local definition as a fractional domain wall,  the disorder operators are actually local observables.

In addition to the  $\mathbb{Z}_2$ symmetry under global spin flips the 2D Ising model has another, more subtle, global symmetry: self-duality \cite{Kramers-1941}.  Duality is a mapping between the low temperature expansion, which is an expansion of the 2D Ising partition function in terms of loops representing the possible configurations of (closed) domain walls, to the high temperature expansion, which is an expansion of the partition function also in terms of loops representing the  extent of spin correlation at high temperatures. Kramers and Wannier showed that these two expansions can be mapped into each other if the coupling constant $K\equiv J/T$ is related to $K^*$, the coupling constant in the dual model, are related by
\begin{equation}
e^{-2K^*}=\tanh K
\label{eq:duality}
\end{equation}
Duality is thus a mapping of the degrees of freedom on the direct lattice at temperature $T$ to the  Ising model on the dual lattice at the dual temperature. It is an exact identity only in the thermodynamic limit.
Under the assumption of a unique phase transition, this relation famously allowed Kramers and Wannier to find the critical temperature of the 2D Ising ferromagnet on  a square lattice at the value obtained (later) by Onsager. 

However, for a finite system Kramers-Wannier duality changes the boundary conditions, e.g. duality maps a system with  periodic boundary conditions to one with  fixed boundary conditions. In this sense, duality is a symmetry of local observables only asymptotically in the thermodynamic limit. The same caveats apply to all the duality mappings that we will discuss below. This issue becomes quite important in three dimensions and in the quantum versions of this problem (even in one-dimension). 

Kadanoff and Ceva realized that duality is actually a local geometric relation. Using this fact they showed that under duality the spin-spin correlation function of the Ising model with coupling constant $K=J/T$ is equal to the correlation function of disorder operators in the  model at dual coupling $K^*$, 
\begin{equation}
\langle \sigma(\bm r) \sigma({\bm r}')\rangle_K=\langle \mu({\tilde {\bm r}}) \mu({\tilde {\bm r}}')\rangle_{K^*}
\label{eq:dual-correlator}
\end{equation}
from which the asymptotic behavior of the correlator of disorder operators follows, consistent with what is expected from Eq.\eqref{eq:disorder-correlator}. Eq.\eqref{eq:dual-correlator} is an identity in the thermodynamic limit. 

\begin{figure}[hbt]
\begin{center}
  \includegraphics[width=0.5\textwidth]{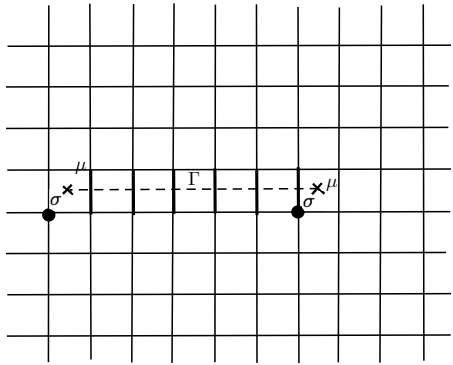}
\caption{Kadanoff-Ceva construction of the Onsager fermion as the product of a spin operator $\sigma$ (the bold black dot) and a disorder operator $\mu$ (the cross at the adjoining dual site).  The disorder operator is equivalent to the insertion of an array of antiferromagnetic bonds on the links pierced by the path $\Gamma$ on the dual lattice.}
\end{center}
\label{fig:onsager_fermion}       
\end{figure}

\subsection{Disorder operators and Onsager fermions}
\label{sec:Onsager-fermions}

It is well known that the 2D classical Ising model is actually a theory of fermions. This was recognized already by Onsager \cite{Onsager-1944} (and extended by Kaufman \cite{Kaufman-1949}) who expressed the solution in terms of a spinor algebra. As a result, the partition function of the 2D classical Ising model is given in terms of the square root of the determinant of a matrix (a Pfaffian). This also implies that the partition function can be expressed as an integral over Grassmann variables  which makes the free-fermion character apparent (see, e.g. Ref. \cite{Fradkin-1979c} and references therein).

Kadanoff and Ceva considered a composite operator made of the product of a spin operator $\sigma({\bm r})$ at lattice site $\bm r$ and a disorder operator at an adjacent dual site ${\tilde {\bm r}}$, such as  shown in Fig. \ref{fig:onsager_fermion}. They denoted this operator by $\psi_\pm(\bm r)$. There  is an ambiguity in this assignment
since one can define two different disorder operators at the  dual sites ${\tilde {\bm r}}_\pm$ adjacent to the site $\bm r$ and, hence, for each site ${\bm r}$, there are two possible such composite operators. It turns out that this is the manifestation of the spinor character of the fermion.

Furthermore, next they  proceeded to consider what happens when one of these composite operators is to  transported around the other along  a closed path of the lattice. 
At the end of this process there will be a closed loop of flipped bonds that will necessarily enclose the spin operator of the other composite operator. 
A closed loop of flipped bonds can be eliminated by a suitable gauge transformation in the interior of the enclosed region, thus restoring the string of flipped bonds to its original ``untangled'' configuration. 
However, the gauge transformation also flips the spin  inside the closed loop, leading to the change of the {\em sign} of the correlator of the composite operators. 
In other words, the composite operator $\psi_\pm({\bm r})$, 
\begin{equation}
\psi_\pm({\bm r}_\pm) \sim \sigma({\bm r}) \mu({\tilde {\bm r}}_\pm)
\label{eq:KC-spinor}
\end{equation}
These operators
behave as  {\em fermions}  which Kadanoff and Ceva identified as with the  fermion (spinor) operator of the Onsager solution of the 2D Ising model \cite{Onsager-1944,Kaufman-1949,Schultz-1964}. 

Kadanoff and Ceva next computed the correlators of these composite operators. I will not discuss here the technical details of the computation. It will suffice to say that they showed that these composite operators obey a set of recursion relations which can be regarded as a discrete linear equation of motion. As noted in Eq.\eqref{eq:KC-spinor} there are two  composite operators of this type and suitable combinations of them can be regarded as a two-component spinor. These algebraic properties enabled them to reduce the computation of the correlation functions of these operators to standard methods.  A key feature of their results was that  correlator changed sign as two such operators were exchanged with each the other and, hence, these operators  are fermions.

They used these results to compute the operator algebra of the non-trivial fixed point of the 2D Ising model. This  was the central aim of that work. 
They showed that the operator algebra of the 2D Ising fixed point  consisted of the fermion operator $\psi$ (with scaling dimension $1/2$), 
the energy density operator $\varepsilon$ (the  relevant thermal operator of the 2D Ising critical point with scaling dimension 1), and the order parameter $\sigma$ 
(with scaling dimension 1/8). By Kramers-Wannier duality, the disorder operator $\mu$ has the same scaling dimension as the order parameter field $\sigma$. 
They also introduced the stress tensor operator $T$, which they showed has scaling dimension 2 (and hence is a marginal operator). 
More than ten years later, Belavin, Polyakov and Zamolodchikov \cite{Belavin-1984} and  Friedan, Qiu and Shenker \cite{Friedan-1984} showed that the operator algebra 
of the 2D Ising model is the simplest example of a non-trivial conformal field theory (CFT) and used this concept  to classify a large class of 2D critical points. 
In this theory the stress-tensor plays a key role as the generator of the Virasoro algebra of the CFT.

\section{Disorder Operators in the 1D Quantum Ising Model}
\label{sec:1D-TFIM}

The equilibrium statistical mechanics of the 2D Ising model is related to the physics of one-dimensional quantum spin systems through the formalism of the transfer matrix (see, e.g. \cite{Schultz-1964}). In this approach, the 2D classical Ising model is viewed as the (discrete) path-integral representation of  the  quantum Ising chain, the Ising model in a transverse magnetic field \cite{Fradkin-1978}, whose Hamiltonian  is
\begin{equation}
H=-\sum_{n=1}^N \sigma_1(n)-\lambda \sum_{n=1}^N \sigma_3(n) \sigma_3(n)
\label{eq:1DTFIM}
\end{equation}
This Hamiltonian commutes with a properly defined  transfer matrix of the classical 2D Ising model. Here $\sigma_1$ and $\sigma_3$ are the two $2 \times 2$ real Pauli matrices and act on the two-dimensional Hilbert space of the spin states at each site $n$ of the chain. This Hamiltonian is solvable for all values of the coupling constant $\lambda$ by means of a Jordan-Wigner transformation to (in this case, Majorana) fermion operators \cite{Schultz-1964,Pfeuty-1970}. 

The Hamiltonian of Eq.\eqref{eq:1DTFIM} is invariant under the global $\mathbb{Z}_2$ symmetry of flipping all spins simultaneously. The operator that effects this symmetry is 
\begin{equation}
Q=Q^{-1}=\prod_{n=1}^N \sigma_1(n)
\label{eq:Z2-Q}
\end{equation} 
This operator commutes with the Hamiltonian, $[Q,H]=0$, and flips the spins,  $Q \sigma_3(n) Q=-\sigma_3(n)$.

\subsection{Disorder operators and kinks}

Just as the classical 2D Ising model, this quantum 1D spin model has two phases: a) the $\lambda < \lambda_c$ disordered phase, in which the $\mathbb{Z}_2$ symmetry is unbroken and $\langle G(\lambda) |\sigma_3(n) |G(\lambda)\rangle=0$, and b) the $\lambda > \lambda_c$ phase in which the $\mathbb{Z}_2$ symmetry is spontaneously broken (in the thermodynamic limit, $N \to \infty$) and $\langle G(\lambda) |\sigma_3(n) |G(\lambda)\rangle \neq 0$. Here 
$|G(\lambda)\rangle$ is the ground state of the Hamiltonian of Eq.\eqref{eq:1DTFIM} at coupling constant $\lambda$. The  phase with unbroken $\mathbb{Z}_2$ symmetry corresponds to the high temperature phase of the 2D classical model and the broken symmetry phase to the low temperature phase of the 2D classical model. The critical coupling $\lambda_c$ corresponds to the Onsager critical temperature. A quantum version of the duality transformation \cite{Fradkin-1978}, under which, up to changes in boundary conditions, the Hamiltonian of Eq.\eqref{eq:1DTFIM} maps onto itself (i.e. it is self-dual) upon the replacement $\lambda \leftrightarrow 1/\lambda$, shows that $\lambda_c=1$, which is the value of exact solution \cite{Pfeuty-1970}.

The  disorder operator in 1D quantum system is the {\em kink} (or {\em domain wall}) creation operator  $\tau_3({\tilde n})$ \cite{Fradkin-1978}
\begin{equation}
\tau_3({\tilde n})=\prod_{j=1}^n \sigma_1(j)
\end{equation}
where $\tilde n$ is the  site of the dual of the  1D lattice, the midpoints between the sites $n$ and $n+1$ of the chain. In the unbroken symmetry phase, $\lambda < \lambda_c$, the disorder operator has an expectation value, $\langle G(\lambda) |\tau_3(n) |G(\lambda)\rangle \neq 0$. Hence,  in the unbroken phase  there  is a condensate  of kinks (or domain walls). Conversely, in the broken symmetry state the disorder operator has zero expectation value since it flips the spins from the boundary to the $n$th site, and  creates a state which us incompatible (orthogonal) with the states of a periodic chain.

\subsection{Majorana fermions}

As it is well known, the 1D quantum Ising model is solved by means of  the Jordan-Wigner transformation that maps a spin chain to a system of fermions
\cite{Lieb-1961,Schultz-1964,Pfeuty-1970}. Thus, we define the operators $\chi_1(n)$ and $\chi_2(n)$
\begin{equation}
\chi_1(n)=\sigma_3(n) \tau_3(n-1), \qquad \chi_2(n)=i \sigma_3(n) \tau_3(n) 
\label{eq:JW}
\end{equation}
These operators are hermitian, $\chi_1(n)^\dagger=\chi_1(n)$ and $\chi_2(n)^\dagger=\chi_2(n)$, and obey the anticommutation rules,
\begin{equation}
\{\chi_1(n),\chi_2(n')\}=0, \qquad \{\chi_1(n),\chi_1(n')\}=\{\chi_2(n),\chi_2(n')\}=2\delta_{n,n'}
\end{equation}
Hence, the operators $\chi_1(n)$ and $\chi_2(n)$ are a set of Majorana (self-adjoint) fermion operators. In this language, the $\mathbb{Z}_2$ symmetry operator $Q$, defined in Eq.\eqref{eq:Z2-Q}, becomes
\begin{equation}
Q=i^N \prod_n\left(\chi_1(n)\chi_2(n)\right)
\label{eq:Z2-Q-majorana}
\end{equation}
The Hamiltonian of Eq.\eqref{eq:1DTFIM} has a quadratic form expressed in terms of  Majorana fermion operators
\begin{equation}
H=-\sum_n i \chi_1(n) \chi_2(n)-\lambda \sum_n i \chi_2(n) \chi_1(n+1)
\label{eq:H-TFIM-Majorana}
\end{equation}
which, of course, is the reason for the solvability (integrability) of the 2D classical Ising model and its 1D quantum cousin.

The Majorana fermions operators of Eq.\eqref{eq:JW} obey (Heisenberg) linear equations of motion
\begin{align}
i \partial_t \chi_1(n)=& i \chi_2(n)-i  \lambda  \chi_2(n-1) \nonumber\\
i \partial_t \chi_2(n)=&-i \chi_1(n)+i  \lambda \chi_1(n+1)
\label{eq:EoM-Majorana}
\end{align}
Near the (quantum) critical point, $\lambda_c=1$, we can take the continuum limit and replace the difference equations of Eq.\eqref{eq:EoM-Majorana} by the differential equations
\begin{align}
i \partial_t \chi_1=& i m \chi_2-i \partial_x  \chi_2 \nonumber\\
i \partial_t \chi_2=&-i m \chi_1+i  \partial_x \chi_1
\label{eq:EoM-Majorana-Dirac}
\end{align}
This is just the Dirac equation in 1+1 space-time dimensions of the Majorana spinor $\chi=(\chi_1,\chi_2)$ with mass $m$ (with lattice spacing $a_0 \to 0$)
\begin{equation}
m=\lim_{\lambda \to 1} \left(\frac{1-\lambda}{a_0\lambda}\right)
\label{eq:scaling-limit}
\end{equation}
holding $m$ fixed. This is the scaling variable  tuning  to the critical point. Therefore, the universality class of the critical point of the 2D classical Ising model (or the quantum critical point of the quantum Ising chain) is  represented by a theory of free relativistic Majorana fermions \cite{Zuber-1977}. At the fixed point this is a CFT with conformal central charge $c=1/2$ \cite{Friedan-1984,Difrancesco-1997}. 

\subsection{Majorana zero modes}

The Hamiltonian of Eq.\eqref{eq:H-TFIM-Majorana} has been the focus of much recent interest, motivated by models of $p$-wave superconducting wires. In this context it is often referred to as the Kitaev chain \cite{Kitaev-2001}. To make this connection clear, it is useful to assign the set of $2N$ Majorana fermion operators, that we denoted by $\chi_1(n)$ and $\chi_2(n)$, to a lattice of $2N$ sites, labeled by $r=1,\ldots,2N$. This new chain can be viewed as the original $N$ sites of the chain and the $N$ sites the dual chain. We now define 
\begin{equation}
\chi(r=2n)=\chi_1(n), \qquad \chi(r=2n+1)=\chi_2(n)
\end{equation}
In this notation the Hamiltonian becomes
\begin{equation}
H=-\sum_{n=1}^{N} i \chi(2n-1) \chi(2n)-\lambda \sum_{n=1}^N i \chi(2n) \chi(2n+1)
\label{eq:H-TFIM-Majorana2}
\end{equation}
Kitaev noted that, if the chain obeys open boundary conditions, in the limit $\lambda \gg 1$, a Majorana chain of this type has exact  ``Majorana zero modes,'' i.e. the boundary operators $\chi(1)$  and $\chi(2N)$ commute with the Hamiltonian. In fact, this property holds  in the entire phase $\lambda >1$. Hence, these so-called Majorana zero modes  persist in this phase up to the critical point, $\lambda=1$, where they disappear as the mass gap closes, $m \to 0$. In contrast,  the opposite phase, $\lambda <1$, all Majorana operators are paired on a  length scale of the order of the correlation length, and an open chain does not have Majorana zero modes \cite{Kitaev-2001}.  

The phase $\lambda>1$ is the broken symmetry phase of the quantum Ising chain. Since in the Ising model the Majorana fermions are closely related with domain walls it is natural that there must be a connection between domain walls and Majorana fermions. Although they are related (and are often confused with each other in the literature) they are different objects. They are related in the sense of the Jordan-Wigner transformation. However, domain walls are (hard-core) bosons whereas Majorana fermions are fermions and, hence, obey different commutation relations. Since in the broken symmetry state, $\lambda>1$, the disorder operator $\tau_3(n)$ does not have an expectation value, the domain walls are not condensed but are, instead, finite energy excitations. In contrast, in this phase the Majorana fermions have a zero energy ``state'' at each end of the chain. As we will see next, the actual state is actually shared by both ends of the chain and has a topological character. This is a manifestation of the fact that the relation between the Majorana chain and the Ising model is two-to-one (and not one-to-ne) since it involves states with different boundary conditions.


Majorana zero modes encode quantum information in a non-local fashion. To see this, we recall that the standard fermion representation of the quantum Ising chain \cite{Lieb-1961,Schultz-1964,Pfeuty-1970} is written in terms of a set of {\em Dirac} fermions operators,
\begin{equation}
c(n)=\chi_1(n)+i \chi_2(n), \quad c^\dagger(n)=\chi_1(n)-i\chi_2(n)
\label{eq:Dirac-fermions}
\end{equation}
which obey canonical anticommutation relations, $\{c(n),c^\dagger(n')\}=\delta_{n.n'}$ and $\{ c(n), c(n')\}=0$. In terms of  Dirac fermions, the Hamiltonian takes the conventional form
\begin{equation}
H=-\sum_n \Big[2 c^\dagger(n)c(n)+\lambda (c^\dagger(n)-c(n))(c^\dagger(n+1)+c(n+1))\Big]
\end{equation}
This Hamiltonian has terms that create and annihilate fermions in pairs and conserve only their parity. It
 has the form of a pairing Hamiltonian familiar from the BCS theory of superconductivity (or the Bogoliubov-de Gennes approximation) \cite{Schrieffer-1964} (normally written in Fourier space), as was recognized already in the work of Schultz, Mattis and Lieb \cite{Schultz-1964}, and in the more recent work of Kitaev \cite{Kitaev-2001}.

For a chain with open boundary conditions all Majorana fermion operators appear in pairs. However, in the $\lambda>1$ phase the Majorana operators $\chi(1)$ and $\chi(2N)$ have no other operators left to pair-up with except each other. Thus, this pair of Majorana operators define a single Dirac operator $c=\chi(1)+i\chi(2N)$ (and its adjoint). This pair of Majorana fermions describes a system with  just two quantum states, the empty state $|0\rangle$ and occupied state $|1\rangle$. This part of the Hilbert space of the chain is not localized anywhere in the chain. In the thermodynamic limit, $N \to \infty$, these two states become degenerate (with zero energy).
Hence, two Majorana zero modes define a single Dirac fermion whose state is either occupied or empty. This non-local encoding of the state is what makes Majorana zero modes interesting  from the point of view of quantum computation \cite{Kitaev-2001,Kitaev-2003}.

Physical systems with Majorana zero modes, in which the properties of the states are encoded non-locally are of great interest as possible platforms for topological Quantum Computing \cite{Dassarma-2007} (TQC). They include such as the fermionic zero modes trapped in the core of two-dimensional $p_x+ip_y$ superconductors \cite{Ivanov-2001} and in the paired states of the fractional quantum Hall fluids \cite{Read-2000}, and in hybrid structures of superconductors  with three-dimensional topological insulators \cite{Fu-2008}. Vortices with Majorana zero modes are non-abelian anyons and encode information in a non-local way, which is why they are possible platforms for TQC.

\section{$\mathbb{Z}_N$ clock models and parafermions}
\label{sec:clock}

\subsection{$\mathbb{Z}_N$ clock models}

We will now consider the case of $\mathbb{Z}_N$ spin systems, also known as clock models (or planar vector Potts models). Clock models, first studied in detail by Jose, Kadanoff, Kirkpatrick and Nelson in 1977 \cite{jose-1977}, are models of spin systems with $N$ states per site. They  are  pictured as a ``clock'' with $N$ hours, and can be regarded as a discrete version of the classical $XY$ (or planar rotor) model, which has a  $O(2)\simeq U(1)$ global symmetry.
 Let us define, at each site ${\bm r}$ of a square lattice, a discrete degree of freedom labelled by $\theta(\bm r)=2\pi n/N$, with $n=0, 1, \ldots, N-1$. We will denote the ``spin'' (the order parameter field) at site $\bm r$ by a complex number of unit modulus, $\sigma(\bm r)=\exp(i \theta (\bm r))$. The partition function of the $\mathbb{Z}_N$ spin model is
\begin{equation}
Z=\sum_{[\theta]} \exp\Big(K \sum_{{\bm r},j=1,2} \cos (\Delta_j \theta({\bm r}))\Big)
\label{eq:ZN-model}
\end{equation}
Here we used the standard notation $\Delta_j \theta(\bm r) \equiv \theta({\bm r}+{\bm e}_j)-\theta(\bm r)$, where ${\bm e}_j$ are the two unit vectors connecting nearest neighboring sites of the square lattice. Here $K=1/T$ is the inverse (dimensionless) temperature. The configuration sum $[\theta]$  runs over the $N$ possible values of all the  discrete angles $[\theta(\bm r)=2\pi n(\bm r)/N]$ at each site of this lattice.
In this case, the system has a global $\mathbb{Z}_N$ symmetry. $\mathbb{Z}_N$ clock models reduce to the Ising model for $N=2$ (two states), to the three-state Potts model for $N=3$  (but not for $N>3$), and to the $XY$ model for $N \to \infty$. 

As in the case of the classical $XY$ model \cite{Villain-1975,jose-1977}, it is be simpler to consider  the closely related (and essentially equivalent) model 
\begin{equation}
Z=\sum_{[\theta],[\ell_j]} \exp\Big(- \sum_{{\bm r},j=1,2} \frac{K}{2}(\Delta_j \theta({\bm r})-2\pi \ell_j(\bm r))^2\Big)
\label{eq:ZN-model-Villain}
\end{equation}
where the degrees of freedom $[\ell_j(\bm r)]$ run over the integers and are defined on the links of the 2D lattice (whose role is to enforce the correct periodicity). For general $N$, an extended model with $\mathbb{Z}_N$ symmetry can be defined \cite{Alcaraz-1980} (including  the $N$ state Potts model which has a permutation symmetry $S_N$).

The $\mathbb{Z}_N$ models share many features with the Ising model. Much as the Ising (and Potts) cousin, $\mathbb{Z}_N$ clock models are self-dual. In the version of the model of Eq.\eqref{eq:ZN-model-Villain}, the dual model has the same form and a dual coupling $\tilde K$ \cite{Dotsenko-1978,Elitzur-1979,Ukawa-1980}.
\begin{equation}
{\tilde K}=\frac{N^2}{4\pi^2K}
\label{eq:ZN-dual-coupling}
\end{equation}
Self-duality then occurs for $K_{SD}=\frac{N}{2\pi}$, or what is the same at the temperature $T_{SD}=\frac{2\pi}{N}$.

The phases (and phase transitions) of the 2D classical $\mathbb{Z}_N$  clock models are well known \cite{jose-1977,Dotsenko-1978,Elitzur-1979,Ukawa-1980,Alcaraz-1980,Savit-1980}. For $N=2,3,4$ the clock model has two phases: a disordered high temperature phase and a low temperature broken symmetry phase (with $N$ degenerate states). 
For $N=2,3,4$ there is a unique continuous phase transition  at the self-dual point. For most $\mathbb{Z}_N$ models, for $N>4$ there is an intermediate critical phase, with continuously varying critical exponents, with critical temperatures $T_{c1} >T_{SD}> T_{c2}$, where $T_{c2}={\tilde T}_{c1}=\frac{4\pi^2}{N^2T_{c1}}$ (the dual temperature of $T_{c1}$).

\subsection{Coulomb gas and the sine-Gordon picture}

Unlike the 2D classical Ising model and the 1D quantum Ising chain, the $\mathbb{Z}_N$ clock models, both in their classical and quantum versions, 
are not integrable systems (except at their self-dual points \cite{Fateev-1982}, or along special values of the parameters of the chiral models \cite{Albertini-1989}), 
and other ways to extract their content must be sought out. 
For $N>4$ the $\mathbb{Z}_N$ clock models are critical for a range of temperatures, as suggested by the duality argument discussed above,  and it is possible to study their critical (and near critical)  
as perturbed Gaussian models \cite{Kadanoff-1979,Fradkin-1980} 
and as generalized Coulomb gases \cite{Kadanoff-1978,Elitzur-1979,Nienhuis-1984}. In the language of conformal field theory, the clock models with $N>4$ have conformal central charge $c=1$ and their fixed point is described by a (compactified) Gaussian (boson) theory.  For $N< 4$ the Gaussian theory is unstable under the RG \cite{jose-1977} and the critical behavior is controlled by a non-trivial fixed point, different for each $N$. The scaling properties of the non-trivial fixed points for $N\leq 4$ were determined by Friedan, Qiu and Shenker  who, in particular, identified the three-state clock (or Potts) model with a minimal model with central charge $c=4/5$ and with a non-trivial operator algebra \cite{Friedan-1984} (and by Dotsenko \cite{Dotsenko-1984}). 

In the Coulomb gas representation $\mathbb{Z}_N$ models are described as a gas magnetic charges (vortices) $m(\bm R)$ and electric charges $n(\bm r)$, where 
$m(\bm R) \in \mathbb{Z}$ and $n(\bm r)\in \mathbb{Z}$. The  partition function of the generalized Coulomb gas is
\begin{equation}
Z=\sum_{\{m(\bm R)\},\{ n(\bm r)\}} \exp(-H[n,m]) \prod_{\bm R} \delta(\sum_{\bm R} m(\bm R)) \prod_{\bm r} \delta(\sum_{\bm r} n(\bm r))
\label{eq:ZN-CG-Z}
\end{equation}
where (at long distances)
\begin{align}
-H[n,m]=&-\frac{N^2}{8K} \sum_{\bm r}n^2(\bm r)-\frac{\pi^2 K}{2} \sum_{\bm R} m^2(\bm R)\nonumber \\
&+\frac{N^2}{4\pi K} \sum_{{\bm r} \neq {\bm r}'} n(\bm r) \ln (|{\bm r}-{\bm r}'|) n({\bm r}')\nonumber\\
&+\pi K \sum_{{\bm R}\neq {\bm R}'} m(\bm R) \ln (|{\bm R}-{\bm R}'|) m({\bm R}')\nonumber\\
&+i N \sum_{{\bm r}, {\bm R}}n(\bm r) \theta({\bm r}-{\bm R}) m(\bm R)
\label{eq:ZN-CG-H}
\end{align}
Here $\theta({\bm r}-{\bm R})$ is the angular position of the magnetic charge at $\bm R$ relative to the electric charge at $\bm r$ (and viceversa). In this picture the $\mathbb{Z}_N$ model is manifestly self-dual, with duality reduced to  exchanging of electric and magnetic charges $n \leftrightarrow m$ (i.e. particles and vortices) and $N^2/(4\pi^2 K) \leftrightarrow  K$, as in Eq.\eqref{eq:ZN-dual-coupling}.

In the dilute gas approximation,  only the electric charges $n=0,\pm 1$ and the magnetic charges $m=0, \pm 1$ effectively contribute to the partition function of Eq. \eqref{eq:ZN-CG-Z}. In this limit, in which the Kosterlitz-Thouless theory applies \cite{Kosterlitz-1973,Kosterlitz-1977,jose-1977}, the partition function of the generalized Coulomb gas can be mapped onto a generalized sine-gordon field theory  of a compactified scalar field $\varphi(\bm x)$ whose (Euclidean) path integral is given by \cite{Wiegmann-1978,Boyanovsky-1989,Lecheminant-2002}
\begin{equation}
Z=\int \mathcal{D} \varphi \; e^{-S(\varphi)}
\end{equation}
where the Euclidean action $S(\varphi)$ is
\begin{equation}
S(\varphi)=\int d^2x \Big[ \frac{1}{2} \left({\bm \nabla} \varphi (\bm x)\right)^2+g \cos\left(\frac{N}{\sqrt{K}} \varphi(\bm x)\right)+{\tilde g} \cos \left(2\pi \sqrt{K} \vartheta(\bm x)\right)\Big]
\label{eq:SG}
\end{equation}
where $ \vartheta$ is the {\em dual field} of the field $\varphi$,
\begin{equation}
i \partial_j \varphi=\epsilon_{jk} \partial_k \vartheta
\label{eq:dual-field}
\end{equation}
and the coupling constants $g$ and $\tilde g$ are related to the fugacities of the generalized Coulomb gas by
\begin{equation}
g \simeq a^{-2}\; e^{-\frac{N^2}{8K}}, \quad {\tilde g}\simeq a^{-2}\; e^{-\frac{\pi^2}{2}K}
\end{equation}
where $a$ is the lattice spacing. In this form duality is the replacement $\varphi \leftrightarrow {\tilde \varphi}$, $K \leftrightarrow N^2/(4\pi^2 K)$ and $g \leftrightarrow {\tilde g}$. Notice that at the self-dual point the coupling constants are equal $g={\tilde g}$. Also, from the action of Eq.\eqref{eq:SG} we can read-off the scaling dimensions of the operator as $\Delta_1=N^2/(4\pi K)$ and $\Delta_2=\pi K$ which imply that for $N > 4$ both operators are irrelevant for values of the stiffness in the range $K_{c1}=\frac{2}{\pi} > K >K_{c2}=\frac{N^2}{8\pi}$ \cite{jose-1977,Elitzur-1979}.

\subsection{The 1D quantum $\mathbb{Z}_N$ model}

It is straightforward to define a 1D $\mathbb{Z}_N$ spin chain,  related to the 2D classical models through the transfer matrix. Aside from their connection with the 2D classical model, these quantum spin chains are physically interesting in their own right \cite{Howes-1983,Mong-2014b}.

To this end we define a set of $N$ quantum states at each site  of the chain, $\{ |2\pi n/N \rangle \}$ (with $n=0,1,\ldots, N-1$). 
These states are eigenstates of the operator $\sigma$, with eigenvalues $\exp(2\pi i n/N)$.
An operator $\tau$  shifts these states downwards by $2\pi/N$, whereas the adjoint operator $\tau^\dagger$ shifts the states upwards by $2\pi/N$.
The operators $\sigma$ and $\tau$ are traceless, $\textrm{tr} \;\tau=\textrm{tr} \; \sigma=0$ (as well as  their adjoints) and, at the same site, satisfy the algebra
\begin{equation}
\tau^N=\sigma^N=I, \quad \tau \sigma=e^{2\pi i/N} \sigma \tau
\label{eq:ZN-algebra}
\end{equation}
but otherwise commute with each other. 
 In terms of these operators, 1D quantum Hamiltonian is
\begin{equation}
H=-\sum_j \left(\tau(j)+\tau^\dagger(j)\right)-\lambda \sum_j \left(\sigma^\dagger(j+1)\sigma(j)+\textrm{h.c.}\right)
\label{eq:H-ZN}
\end{equation}

The classical and quantum $\mathbb{Z}_N$ models have been generalized by several authors \cite{Elitzur-1979,Alcaraz-1980} to a class of models which largely have the same structure of 
phase transitions. An interesting generalization are the case of the chiral Potts and $\mathbb{Z}_N$ models \cite{Ostlund-1981,Huse-1983,Howes-1983,VonGehlen-1985,Fendley-2012}. The chiral Potts model has an interesting phase structure, 
and for some values of the chiral parameter is integrable. In its quantum version, the simplest Hamiltonian of the chiral $\mathbb{Z}_N$ chain is
\begin{equation}
H=-\sum_j \left(e^{i\theta} \tau(j)+\textrm{h.c.}\right)-\lambda \sum_j \left(e^{i\phi} \sigma^\dagger(j+1)\sigma(j)+\textrm{h.c.}\right)
\label{eq:H-chiral-ZN}
\end{equation}
where $\phi$ and $\theta$ are two parameters with periodicity $2\pi$. The integrable version of the Hamiltonian of the chiral Potts model of Eq.\eqref{eq:H-chiral-ZN} has additional terms with increasing powers of the $\sigma$ and $\tau$ operators with fine-tuned coupling constants.

\subsection{$\mathbb{Z}_N$ disorder operators}

The $\mathbb{Z}_N$ models have a discrete symmetry and an $N$-fold degenerate broken symmetry phase. 
In the low temperature phase the $N-1$ possible order parameters, the $\mathbb{Z}_N$ spins $\sigma_n({\bm r})=\exp(i 2\pi n({\bm r})/N)$ 
(with $n=1,\ldots,N$), have a non-vanishing expectation value. For this reason the  $\mathbb{Z}_N$ clock models have $N-1$ types of domain walls, 
closed paths on the dual lattice that separate regions with different broken symmetry states differing by an angle $2\pi n/N$ (with $n=1,\ldots,N-1$). 

Likewise, the 2D classical $\mathbb{Z}_N$ clock models have $N-1$ types of disorder operators, that can be defined by  analogy with the Kadanoff-Ceva construction used in the Ising model. 
In the $\mathbb{Z}_N$ case there are $N-1$ types of disorder operators,  which are denoted by $\mu_m(\bm R)$. 
These operators create a fractional domain wall at which the $\mathbb{Z}_N$ spins rotate by an angle $2\pi m/N$ 
along a path on the dual lattice ending at the plaquette labeled by the dual site ${\bm R}$ (here I am using the same geometry shown in Fig.\ref{fig:disorder-operators}). 

In the case of the $\mathbb{Z}_N$ clock models the disorder operators can be represented by a the coupling of the $\mathbb{Z}_N$ spins to a background discrete gauge field
$A_j(\bm r)=2\pi m_j(\bm r)/N$  defined on the links $({\bm r}, {\bm r}+{\bm e}_j)$ (with $j=1,2$) of the square lattice. 
Here we are using the language of lattice gauge theory \cite{Wilson-1974,Kogut-1979} which is particularly useful in this context \cite{Fradkin-1978b}. 
This amounts to changing the energy functional of the $\mathbb{Z}_N$ model, Eq.\eqref{eq:ZN-model} and Eq.\eqref{eq:ZN-model-Villain}, by the minimal coupling prescription, 
i.e. to make the replacement $\Delta_j \theta(\bm r) \mapsto \Delta_j \theta(\bm r)-A_j(\bm r)$ in every term of the energy functional. 
Let $\Phi(\bm R)=\epsilon_{ij} \Delta_i A_j(\bm r)$ be the flux (or circulation) of the gauge field $A_j(\bm r)$ around the plaquette centered at the dual site $\bm R$. 

The disorder operator (or frustration) $\mu_q(\bm R)$ of charge $q$ at the plaquette of dual site $\bm R$ represents the insertion of a flux $2\pi q/N$ of the gauge field $A_j(\bm r)$ 
at the plaquette labeled by the dual lattice site $\bm R$.
Thus, a disorder operator can then  be viewed as a {\em magnetic charge} of flux $2\pi q/N$ at that plaquette. In this picture, the spin degrees of freedom play the role of a (fluctuating) matter field 
and the gauge field plays the role of a fixed (or background) gauge field. For this reason, a spin degree of freedom $\sigma(\bm r)=\exp(2\pi n(\bm r)/N)$ carries the unit ``electric charge" 
and $\sigma(\bm r)^p$ carries $p$ units of ``electric charge''. 
In this picture, the Kadanoff-Ceva construction of the disorder operator corresponds to choosing the (axial)  gauge in which the gauge fields are equal to $2\pi q/N$ 
along the bonds (links) of the lattice pierced by the path on the dual  lattice shown in Fig.\ref{fig:disorder-operators}. In the Coulomb gas representation of Eq.\eqref{eq:ZN-CG-Z} and \eqref{eq:ZN-CG-H}, a disorder operator of magnetic charge $q/N$ at dual site $\bm R$ amounts to  shifting the vortex charge at $\bm R$ my a fractional amount, $m(\bm R) \to m(\bm R)+q/N$. Likewise, the insertion of a spin operator of charge $p$ at site $\bm r$ amounts to a shift of the charge variables of the generalized Coulomb gas by $n(\bm r) \to n(\bm r) + p/N$.

From this construction, it would seem that here too the disorder operators are inherently highly non-local.
However, just as in the case of the 2D Ising model, the disorder operators have exponentially decaying correlation functions in the ordered phase and hence behave as local operators. 
In this case too, disorder operators have a non-zero expectation value in 
the disordered phase and their connected correlation functions decay exponentially with distance in the ordered phase. 
As in the case of the Ising model, order and disorder operators map into each other under the duality transformation.

In the case of the $\mathbb{Z}_N$ quantum chains, whose Hamiltonian is given by Eq.\eqref{eq:H-ZN}, the disorder operators are  operators that create kinks. In this case, 
the kink operators  rotate the $\mathbb{Z}_N$ spins by an angle of $2\pi /N$. The disorder operators of the quantum $\mathbb{Z}_N$ chain are
\begin{equation}
{\tilde \sigma}^\dagger(j)=\prod_{k<j} {\tau^\dagger}(k)
\label{eq:sigma-tilde}
\end{equation}
Similarly, we can define the operators ${\tilde R}^\dagger$ on the dual lattice of the 1D chain
\begin{equation}
{\tilde \tau}^\dagger(j)=\sigma^\dagger(j-1)\sigma(j)
\label{eq:tau-tilde}
\end{equation}
It is easy to see that the dual operators ${\tilde \sigma}$ and ${\tilde \tau}$ satisfy the same algebra of Eq.\eqref{eq:ZN-algebra}. Up to subtleties related to boundary terms, the Hamiltonian of Eq.\eqref{eq:H-ZN} has the same form, i.e. it is self-dual, under this duality transformation upon the replacement of the coupling constant $\lambda \mapsto {\tilde \lambda}=1/\lambda$. Here too, duality is a symmetry up to boundary terms in the Hamiltonian (and changes of boundary conditions).

It is a simple excercise to see that the disorder operators, $\tau(j)$, have a non-zero expectation value of the ground state of the quantum $\mathbb{Z}_N$ chain in its disordered phase 
(for $\lambda$ smaller than a critical value $\lambda_{c1}$) and have exponentially decaying correlations in the ordered phase 
(with $\lambda$ larger than $\lambda_{c2}> \lambda_{c1}$). Order and disordered phases map into each other under duality.

\subsection{Parafermions}

The generalization of the Onsager fermions of the Ising model to the $\mathbb{Z}_N$ clock model are known as {\em parafermions} and were introduced Kadanoff and Fradkin \cite{Fradkin-1980}. 
Parafermion operators are charge-flux composite operators that can be defined both in the 2D classical model and in the related quantum chains. 
These charge-flux composites share many properties of  {\em anyon} operators of quantum systems in $2+1$ dimensions \cite{Wilczek-1982}.
Anyons play a key role in the theory of fractional quantum Hall fluids \cite{Laughlin-1983,Halperin-1984,Zhang-1989,Lopez-1991,Wen-1995}, 
and are closely related (and  partially inspired by) `t Hooft's concept of oblique confinement,  originally proposed in the context of gauge theory \cite{thooft-1979}.

Parafermions arise in  $\mathbb{Z}_N$ quantum chains as a straightforward generalization of the Jordan-Wigner transformation. 
We  define the parafermion operator $\psi_{p,q}(j)$ as a product of a disorder operator of charge $q/N$ and an order operator of charge $p$
\begin{equation}
\psi_{p,q}(j)=\prod_{k<j} (\tau(k))^q \sigma_p(j)
\label{eq:JW-ZN}
\end{equation}
It easy to see that these operators satisfy
\begin{equation}
(\psi_{p,q}(j))^N=I, \quad \psi_{p,q}(j) \psi_{p',q'}(j')=e^{i \frac{2\pi}{N} (pq'+qp')} \psi_{p',q'}(j') \psi_{p,q}(j)
\label{eq:parafermion-commutation}
\end{equation}
It is also easy to see that the 1D quantum Hamiltonian of Eq.\eqref{eq:H-ZN} can be rewritten as a bilinear form of parafermions. 
However, due to the non-canonical form of the parafermion commutation relations, Eq.\eqref{eq:parafermion-commutation}, 
the equations of motion of parafermions are not linear and this Hamiltonian is not integrable (except at the self-dual point). 

Parafermions were introduced in Ref. \cite{Fradkin-1980} in the context of the classical 2D $\mathbb{Z}_N$ models as composite operators of 
order operators $\sigma_p(\bm r)$ and disorder operators $\mu_q(\bm R)$. 
There, just as in the case of the 2D Ising model, the order and disorder operators are mutually non-local to each other and that taking an order operator on a 
path that contains the order operator yields a phase change by $\exp(\pm i 2\pi pq/N)$, where the sign of the exponent depends on the orientation of the path. 

Furthermore, upon defining the parafermion operator resulting from the {\em fusion} of the order operator $\sigma_p$ and the disorder operator $\mu_q$ 
using the operator product expansion \cite{Kadanoff-1969,Wilson-1969,Polyakov-1970}, we showed that the correlation function of two such operators, 
denoted by $\psi_{p,q}({\bm r}_1)$ and $\psi_{-p,-q}({\bm r}_2)$, changes by the same  phase factor as one composite operator circles the other. 
Explicit results for the parafermion correlation functions were obtained in the critical regime, by relating the $\mathbb{Z}_N$ model to the gaussian model, with the result
\begin{equation}
\langle \psi_{p,q}({\bm r}_1) \psi_{-p,-q}({\bm r}_2)\rangle=\frac{\exp[-2pqi \theta/N]}{|{\bm r}_1-{\bm r}_2|^{2\Delta_{p,q}}}
\label{eq:parafermion-correlator}
\end{equation}
where $\theta$ is the angle from ${\bm r}_2$ measured from ${\bm r}_1$ (i.e. there is a branch cut from each disorder operator running along the negative $x$ axis). 
This result implies that the parafermion operator creates a state with intrinsic (fractional) angular momentum $pq/N$. In Eq.\eqref{eq:parafermion-correlator}
 $\Delta_{p,q}$ is the scaling dimension of the parafermion operator which is given by
\begin{equation}
\Delta_{p,q}=\frac{p^2}{2\pi K}+\frac{2\pi q^2}{N^2} K
\label{eq:scaling-dimension}
\end{equation}
The existence of parafermion operators in 2D classical (and 1D quantum) $\mathbb{Z}_N$ models was confirmed by Dotsenko \cite{Dotsenko-1984} and by Zamolodchikov and Fateev \cite{Zamolodchikov-1985,Zamolodchikov-1987} 
who showed that the conformal field theory of the $\mathbb{Z}_3$ clock model \cite{Difrancesco-1997} (describing its critical point) has an operator of scaling dimension $1/3$ 
and conformal spin $1/3$, consistent with the predictions of Ref.\cite{Fradkin-1980}.

\subsection{Parafermion zero modes}

The search for platforms for non-abelian topological quantum computing has recently focused on hybrid structures of different quantum Hall states \cite{Clarke-2013}, and between quantum Hall states and superconductors \cite{Lindner-2012,Mong-2014,Vaezi-2014}. These novel platforms, whose experimental realizations is currently an area of intense research, define one-dimensional channels confined by these hybrid structures which, as it turns out, harbor parafermion zero modes \cite{Fendley-2012,Alicea-2016}. These proposals have brought renewed interest in the physics of parafermions \cite{Fendley-2012,Mong-2014,Alicea-2016} and of parafermion zero modes \cite{Fendley-2012}.

In their simplest physical realization the models involve a line junction of two quantum Hall states with a conventional charge $2e$ superconductor occupying a central segment of the junction \cite{Clarke-2013,Mong-2014}. These authors showed that in this system there are $\mathbb{Z}_3$ parafermion zero modes trapped at the endpoints of the superconducting wire in the junction. An array of such wires will then allow for these parafermions to be fused and braided. What is important in this context is that the fusion and braiding properties of these parafermion zero modes is described by the fusion algebra of the  $\mathbb{Z}_3$ (or in general, $\mathbb{Z}_N$) conformal field theory which involves non-abelian fractional statistics.

The conceptually simplest version of this scheme was proposed by Clarke, Alicea and Shtengel \cite{Clarke-2013}.
They considered two fractional quantum Hall fluids, each at at the Laughlin filling fraction $\nu=1/m$ (with $m$ an odd integer) 
but with opposite spin polarizations. This effect that may be achieved by tuning the gyromagnetic factor $g$ from positive to negative accross the line junction. 
The edges of the two fluids form a line junction described by two counter-propagating edge states. 
They further assumed that the outer section of the region comprised between the two quantum Hall fluids   is occupied by a superconductor (with high critical field), 
with pairing field $\Delta_{sc}$, and the remaining region by an insulator (with strong spin orbit coupling), where an gap $\mathcal{M}$ on the edge state spectrum 
opens due to backscattering processes of electrons between the two edge states. 

The  two counter-propagating edge states are described by a Bose field $\varphi(x,t)$ and its canonically conjugate momentum $\Pi(x,t)=\partial_x \vartheta$, 
where $\vartheta(x,t)$ is the dual field of $\varphi(x,t)$ \cite{Wen-1995,Fradkin-2013}. These fields obey the commutation relations $[\varphi(x),\vartheta(y)]=i\frac{\pi}{m} \Theta(x-y)$, with $\Theta(x)$ being the Heaviside step function. The effective quantum Hamiltonian density of the line junction is
\begin{equation}
\mathcal{H}=\frac{m v}{2\pi} \Big[ (\partial_x \vartheta (x))^2+(\partial_x \varphi(x))^2\Big]- \Delta_{sc}(x) \cos(2m\varphi(x))-\mathcal{M}(x) \cos(2m\vartheta(x))
\label{eq:H-qh-junction}
\end{equation}
where $v$ is the speed of the edge modes, and where we have allowed for the superconducting gap $\Delta_{sc}$ and the backscattering gap 
$\mathcal{M}$ to be position-dependent (reflecting the geometry of the junction).

Remarkably, Eq.\eqref{eq:H-qh-junction} is the Hamiltonian of a one-dimensional system whose two-dimensional Euclidean action is, up to a simple rescaling of the fields, 
given by the action of the $\mathbb{Z}_N$ model of Eq.\eqref{eq:SG}, for  $N=2m$. 
Notice that in this representation duality is simply the replacement $\varphi \leftrightarrow \vartheta$ together with $\Delta_{sc} \leftrightarrow \mathcal{M}$ 
(i.e. swapping the superconducting and insulating regions).
More general constructions, which allowed for the realization of $\mathbb{Z}_N$ models with general (even and odd)  values of $N$ 
have also been proposed \cite{Lindner-2012,Clarke-2014,Mong-2014,Vaezi-2014,Alicea-2016,Alexandradinata-2016}. 
In the language of the Hamiltonian of Eq.\eqref{eq:H-qh-junction} the junction with the geometry  considered in Ref. \cite{Clarke-2013} is described as two domain walls between the superconducting regions (where $\Delta_{sc} \neq 0$) 
and the insulating region (where $\mathcal{M} \neq 0$). The important result of Ref.\cite{Clarke-2013} is that the domain walls trap parafermion zero modes, and that these junctions behave as non-abelian anyons that can be used for topological quantum computation \cite{Dassarma-2007}. 

\section{Topological spin chains}
\label{sec:spin-chains}

We will now discuss the role of disorder operators in  one-dimensional quantum spin-$S$ antiferromagnetic quantumHeisenberg models. The Hamiltonian for a chain with $N$ sites is
\begin{equation}
H=J\sum_{n=1}^N {\bm S}(n)\cdot {\bm S}(n+1) 
\label{eq:H-QAF-S}
\end{equation}
Here ${\bm S}$ are spin $S$ operators,  with $S$ being either an integer or a half-integer. 

The ground state and low lying spectrum of spin-$S$ Heisenberg antiferromagnets is well understood. 
A fundamental result by Haldane \cite{Haldane-1983a,Haldane-1983b} shows that integer and half integer spin chains behave quite differently. 
It has long been known that the $S=1/2$ spin chain, which is exactly solvable by Bethe ansatz methods \cite{Bethe-1931,Yang-1966}, 
has a gapless spectrum and its low-energy behavior is described by a an $SU(2)_1$ conformal field theory \cite{Affleck-1986,Affleck-1987}. 

Using semi-classical methods Haldane showed that spin-$S$ quantum Heisenberg antiferromagnetic chains are described by an effective field theory of the form of a 
non-linear sigma model with a topological $\theta$ term. The Euclidean (imaginary time) path integral  of this non-linear sigma model (NLSM) is
\begin{equation}
Z_{\rm NLSM}=\int \mathcal{D} {\bm n} \prod_{\bm x} \delta({\bm n}(\bm x)^2-1) \exp\left(-\int d^2x \mathcal{L}[\bm n]\right)
\end{equation}
where  the field ${\bm n}(\bm x)$,  representing the slowly varying components of the N\'eel order parameter of the spin chain, is a three-component real unit vector field,
satisfying the local constraint ${\bm n}(\bm x)^2=1$. The Lagrangian density $\mathcal{L}[\bm n]$ is
\begin{equation}
\mathcal{L}[\bm n]=\frac{1}{2g}\Big[\frac{1}{v_s} \left(\partial_t {\bm n}(\bm x)\right)^2+v_s\left(\partial_x {\bm n}(\bm x)\right)^2\Big]
+i\frac{\theta}{8\pi}\epsilon_{ij}{\bm n}(\bm x)\cdot \partial_i {\bm n}(\bm x)\times \partial_j{\bm n}(\bm x)
\label{eq:nlsm-S}
\end{equation}
The last term in Eq.\eqref{eq:nlsm-S} is known as the $\theta$-term and,as we will see below, governs the topological character of the states.
In Eq.\eqref{eq:nlsm-S} the dimensionless coupling constant is $g=2/S$, $v_s\simeq 2JS a_0$ is the (non-universal) spin wave velocity, and $\theta=2\pi S$. 
A detailed derivation can be found in Ref.\cite{Fradkin-2013}. 

Haldane's result follows from two observations. One is Polyakov's result that the NLSM in two space-time dimensions is asymptotically free 
and that in the infrared the coupling constant $g$ flows to strong coupling under the renormalization group \cite{Polyakov-1975b}. 
The second observation is that the quantity $\mathcal{Q}[\bm n]$
\begin{equation}
\mathcal{Q}[\bm n]=\frac{1}{8\pi} \int d^2x\; \epsilon_{ij}\; {\bm n}(\bm x)\cdot \partial_i {\bm n}(\bm x)\times \partial_j{\bm n}(\bm x)
\label{eq:top-charge}
\end{equation}
is a topological invariant, known as the topological charge or winding number.
 The topological charge $\mathcal{Q}$ takes integer values that classify the field configurations into the homotopy classes of  maps of the two-dimensional space 
 compactified to the two-sphere $S_2$ onto the target space two-sphere $S_2$ of the configuration space of the order parameter field ${\bm n}$, 
 i.e. the homotopy group $\pi_2(S_2)=\mathbb{Z}$. 
 Therefore, the second term in the Lagrangian of Eq \eqref{eq:nlsm-S} is a topological term that contributes to the Euclidean action by 
 the amount $S_{\rm topo}=i \theta \mathcal{Q}$. Since $\theta=2\pi S$, it follows that the topological weight of a configuration ${\bm n}(\bm x)$ to the path integral is 
 \begin{equation}
 e^{i\theta \mathcal{Q}[\bm n]}=(-1)^{2\pi S\mathcal{Q}[\bm n]}=
 \begin{cases}
\;\;\, 1, \quad {\rm for} \; S\in \mathbb{Z}\\
 -1,\quad {\rm for} \; S\in \mathbb{Z}+\frac{1}{2}
 \end{cases}
 \end{equation}
 Hence, for all integer values of the spin $S$, the topological weight of a configuration is $+1$,
and  for half-integer values of the spin $S$ the topological weight is $(-1)^{\mathcal{Q}}$. 

This analysis implies that antiferromagnetic 
 Heisenberg spin chains with half-integer spin have the same behavior as for $S=1/2$ and are gapless. Instead, for all integer spin they are gapped since the NLSM 
 (without a topological term) is always in a massive phase for all values of the coupling constant \cite{Haldane-1983a,Haldane-1983b},

 We will now focus on the integer spin chains. This result implies that the integer spin $S$ chains have a ground state without long range order and exhibit a Haldane spin gap in the low energy spectrum \cite{Haldane-1983a}. 
 We will see that for integer spin chains there is an analog of the disorder operator 
 that plays a key role in the case of {\em integer} antiferromagnetic quantum spin chains. 
 The analog of the disorder operator for the integer spin chains is the {\em string 
 operator},  introduced by den Nijs and Rommelse \cite{den-Nijs-1989} in the context of the problem of pre-roughening transitions of classical crystal surfaces. It is
 related to the $S=1$ spin chain 
 by the transfer matrix construction. 
 The equal-time correlator of the string operator is \cite{den-Nijs-1989}
 \begin{equation}
 G_{\rm string}[n]=\langle 0 \vert  \prod_{a=x,y,z} S_a(m) \exp\left(i \pi \sum_{k=m}^{m+n} S_a(k)\right) S_a(m+n) \vert 0\rangle
 \label{eq:string}
 \end{equation}
where $\vert 0 \rangle$ is the ground state of the $S=1$ chain. den Nijs and Rommelse showed that this operator has a non-zero expectation value in the 
Haldane phase, and exhibits  exponential decay with distance in the dimerized (or ``valence bond crystal'') phase of a $S=1$ 
Heisenberg antiferromagnetic chain whose Hamiltonian also includes a large enough biquadratic exchange term, $K \sum_n ({\bm S}(n) \cdot {\bm S}(n+1))^2$. 
Hence, the string operator has a non-vanishing expectation value in the ground state of Haldane phase of the spin $S=1$ chain, which is a state which does not break any symmetries, 
and, in this sense, it is a disordered phase. 
In contrast, the string operator has a vanishing expectation value in the dimerized phase,  a spin-singlet ground state that breaks spontaneously translation symmetry. 
These results were subsequently investigated in detail numerically by Girvin and Arovas \cite{Girvin-1989}, and  extended  to  spin chains with arbitrary integer spin 
$S$ by Tasaki \cite{Tasaki-1991} and by Oshikawa \cite{Oshikawa-1992}.

We close this discussion by noting that the ground states of the spin $S$ chain in the Haldane phase of odd and even integer spin chains actually are  not equivalent. 
Even though both types of antiferromagnetic chains have a Haldane gap, the odd integer spin chains are actually in a topological phase while the even integer spin
 chains are not. Indeed, it has been known for quite some time that an $S=1$ antiferromagnetic Heisenberg spin chain with $N$  
 sites and open boundary conditions behaves as it it had two spin-1/2 degrees of freedom, one at each end of the chain \cite{Hagiwara-1990,Kennedy-1990}. 
 These spin-1/2 ``edge states'' are present provided the spin symmetry of the chain is exact.  
 In contrast, for $S=2$ spin chains these edge degrees of freedom 
 form local spin-singlets and are gapped. For these reason the Haldane state for spin chains with spin $S$ odd are examples of symmetry-protected topological phases \cite{Senthil-2015} while the 
 spin chains with $S$ even are topologically trivial. This same behavior is also seen long spin $S=1$ chains with periodic boundary conditions which have  a double degeneracy 
 of the entanglement spectrum of a finite segment \cite{Pollmann-2010}. 
 
 These ``dangling'' spin-1/2 degrees of freedom of long but finite open spin chains 
 (with odd integer spin $S$) are analogs of the Majorana fermions of the Ising chains and the parafermions of the $\mathbb{Z}_N$ spin models  discussed in the preceding sections. 
 Intuitively, one can picture the local $S=1$ degrees of freedom of the chain as each being made of two $S=1/2$ spins (projected onto the triplet manifold). 
 Then, the local $S=1/2$ edge states are fractionalized $S=1$ degrees of freedom  whose missing spin-1/2 partner resides at the other end of the chain \cite{Kennedy-1990}. 
 These behaviors cannot occur in ground states with only short range entanglement and can only happen  in a topological phase.

\section{Disorder operators, confinement and topological phases of matter}
\label{sec:IGT}

We now turn to the role of disorder operators in higher dimensional systems.  Here we consider the specific case of the 3D Ising model 
and its Kramers-Wannier dual, the 3D Ising gauge theory \cite{Wegner-1971,Balian-1975,Wilson-1974,Fradkin-1978,Kogut-1979}.  

\subsection{Disorder operators in 3D}

In the case of the 3D Ising model the natural generalization of the Kadanoff-Ceva construction to 3D is again a fractional domain wall \cite{Fradkin-1978b}. 
However in 3D domain walls are closed surfaces and fractional domain walls are surfaces with a 1D closed boundary. 
The same applies to the case of the 3D $\mathbb{Z}_N$ spin system (and, for the matter, for any 3D spin system with a discrete symmetry) which, for brevity,  
we will not discuss here.  
As in the 2D case, a fractional domain wall is created by an operator that flips the signs 
of the Ising couplings from ferro to antiferromagnetic on a set of bonds of the 3D lattice normal to a surface $\Sigma$ whose boundary is $\Gamma$.  

Paraphrasing the 2D construction of Eq.\eqref{eq:mumu}, we can define the ratio of the partition functions with and without the operator that creates the fractional domain wall to be the expectation value of the disorder operator which now is 
\begin{equation}
\frac{Z[\Sigma]}{Z}=\exp(-\Delta F[\Sigma]/T)
\label{eq:disorder-3D}
\end{equation}
In the 3D Ising model, $T<T_c$, the free energy cost $\Delta F[\Sigma]$ of  a fractional domain wall 
scales with the area $\mathcal{A}[\Sigma]$ of the surface of the fractional wall, $\Delta F[\Sigma] = \rho(T) \mathcal{A}[\Sigma]$, 
where $\rho(T)$ is the surface tension of the domain wall, a quantity that is finite in the ordered phase. Conversely, in the high temperature, $T>T_c$, 
disordered phase, the free energy cost scales with the perimeter $\mathcal{P}$ of the boundary $\Gamma=\partial \Sigma$ of the fractional domain wall, 
$\Delta F[\Sigma]={\bar \rho} (T) \mathcal{P}[\Gamma]$.
Hence, we find that for asymptotically large surfaces $\Sigma$, the ratio of partition functions behaves as
\begin{equation}
\frac{Z[\Sigma]}{Z}\propto
\begin{cases}
\exp\left(-\frac{\rho(T)}{T} \mathcal{A}[\Sigma]\right), \quad T <T_c\\
\exp\left(-\frac{{\bar \rho}(T)}{T} \mathcal{P}[\Gamma]\right), \quad T>T_c
\end{cases}
\label{eq:disorder-3D-behavior}
\end{equation}
Thus, in the 3D Ising model (and on all  models with a global discrete symmetry) the disorder operators are non-local operators.

\subsection{Monopole condensate and confinement}

Disorder operators play an important role  in understanding the phases of gauge theory. For simplicity and brevity here we will discuss only the case of the 3D Ising gauge theory but these concepts can (and have) been extended to other cases.
 
 The partition function of the 3D Ising gauge theory is \cite{Wegner-1971,Balian-1975}
 \begin{equation}
 Z_{\rm gauge}=\sum_{[\sigma_j(\bm r)]} \exp\left(K^*\sum_{{\bm r}, j,k=1,2,3} \sigma_j(\bm r) \sigma_k({\bm r}+{\bm e}_j)\sigma_k(\bm r) \sigma_j({\bm r}+{\bm e}_k)\right)
 \label{eq:Z2gauge}
 \end{equation}
where the sum runs over the configurations of Ising degrees of freedom $\sigma_j(\bm r)=\pm 1$  on the links of the 3D cubic lattice, and $K^*$ is the coupling constant. 
The sum in the (Euclidean) action of Eq. \eqref{eq:Z2gauge} runs over the plaquettes of the cubic lattice, labelled by the sites ${\bm r}$ and pairs of directions $ j$ and $k$. 
The action of Eq.\eqref{eq:Z2gauge} is invariant under  arbitrary  local $\mathbb{Z}_2$ 
gauge transformations $\sigma_j(\bm r) \mapsto s(\bm b) \sigma_j(\bm r) s({\bm r}+{\bm e}_j)$, with $s(\bm r)=\pm 1$, at every site $\bm r$ of the 3D lattice. 
The existence of a local gauge invariance  requires that the only observables with a  non-zero expectation value to be gauge invariant \cite{Elitzur-1975}. 
A generic gauge-invariant observable is a Wilson loop operator \cite{Wilson-1974} defined on a closed loop $\gamma$ of the cubic lattice,
\begin{equation}
W[\gamma]=\prod_{({\bm r},{\bm r}+{\bm e}_j)\in \gamma} \sigma_j(\bm r)
\label{eq:Wilson-loop}
\end{equation}
where $({\bm r},{\bm r}+{\bm e}_j)$ denotes the set of links on the closed loop $\gamma$.

The 3D Ising gauge theory is the Kramers-Wannier dual of the 3D Ising model \cite{Wegner-1971}. Under duality the coupling constant $K^*$ is 
related to the coupling constant $K$ 
of the (dual) 3D Ising model by the  relation of Eq.\eqref{eq:duality}. 
A direct consequence of  duality, is that the dual of the Wilson loop operator of Eq.\eqref{eq:Wilson-loop} 
 is  the fractional domain wall of Eq.\eqref{eq:disorder-3D}. Under duality the ordered phase of the 3D Ising model ($T<T_c$, or equivalently $K>K_c$) 
 maps onto the confining phase of the gauge theory, $K^*<K_c^*$. In this phase the expectation value of the Wilson loop operator satisfies the area law, which is
 Wilson's criterion for confinement \cite{Wilson-1974}), and is consistent with Eq.\eqref{eq:disorder-3D-behavior}.
  Likewise, the disordered phase if the Ising model maps onto the deconfined phase of the gauge theory, where the perimeter law of Eq.\eqref{eq:disorder-3D-behavior} holds. Here too, we must note that duality has a subtle effect on boundary conditions. As we will see, this particularly important in the deconfined phase of the gauge theory.

On the other hand, the 3D Ising model has a local order parameter, the local magnetization, which has a non-vanishing expectation value in the ordered phase. 
Under duality, the local magnetization at site $\bm r$ is identified as a ``monopole'' operator: 
the sign of the coupling constant $K^*$ is changed from positive to negative on a tube of plaquettes pierced by some path ${\tilde \Gamma}[\bm r]$ of the dual lattice 
ending at the cube dual to the site $\bm r$. Such an operator favors the creation of a Dirac string with flux $\pi$ on each plaquette in the tube. 
This operator has a non-vanishing expectation value in the confining phase of the gauge theory which then may  be regarded as a condensate 
of $\pi$ fluxes \cite{Fradkin-1978}. This operator is the analog of the Kadanoff-Ceva disorder operator for the gauge theory.
In this sense, confining phases of gauge theory are viewed as   condensates  of magnetic monopoles \cite{Polyakov-1975,thooft-1979,Seiberg-1994}. 
In contrast, in the deconfined phase (which is dual to the disordered phase of the Ising ferromagnet) this operator has a vanishing expectation value, 
and its correlation functions decay exponentially with distance. 

\subsection{Quantum Hamiltonian picture of duality}

More insight may be gained by looking at the quantum Hamiltonian associated with the Ising gauge theory in 2+1 dimensions (related to the 3D Ising gauge theory 
through the transfer matrix) \cite{Kogut-1979}
\begin{equation}
H=-\sum_{{\bm r}, j=1,2} \sigma^1_j(\bm r)-\lambda \sum_{\bm r} \sigma^3_1(\bm r) \sigma^3_2({\bm r}+{\bm e}_1)\sigma^3_2(\bm r) \sigma^3_1({\bm r}+{\bm e}_2)
\label{eq:H-IGT}
\end{equation}
where $\sigma^1_j(\bm r)$ and $\sigma^3_j(\bm r)$ are Pauli matrices defined on the links of the square lattice. The Hilbert space of gauge-invariant states, denoted by  $\vert \textrm{Phys}\rangle$,  is the vector space of states that obey the Gauss law constraint,
\begin{equation}
\sigma^1_1(\bm r) \sigma^1_1({\bm r}-{\bm e}_1)\sigma^1_2(\bm r) \sigma^1_2({\bm r}-{\bm e}_2)\vert \textrm{Phys}\rangle =\vert \textrm{Phys}\rangle
\label{eq:gauss}
\end{equation}
at each site $\bm r$ of the lattice. The operator on the left hand side of Eq.\eqref{eq:gauss} is the generator of local $\mathbb{Z}_2$ gauge transformations.

This quantum Hamiltonian describes a systems with the same two phases discussed above: a confining phase for $\lambda<\lambda_c$, 
and a deconfined phase for $\lambda>\lambda_c$. In addition for its interest as a the simplest gauge theory, and as the dual of the 2D quantum Ising model, the $\mathbb{Z}_2$  Ising gauge theory plays an important role in the theory of $\mathbb{Z}_2$ spin liquids of frustrated quantum antiferromagnets \cite{Read-1991} and the related quantum dimer models \cite{Rokhsar-1988,Moessner-2001,Moessner-2001c,Fradkin-2013}.

In this language the monopole operator is
\begin{equation}
\tau_3({\tilde {\bm r}})=\prod_{\ell \in \Gamma[{\tilde {\bm r}}]} \sigma^1(\ell)
\end{equation}
where $\ell$ denotes the set of links pierced by the path $\Gamma$ on the dual lattice ending at the plaquette ${\tilde {\bm r}}$ (with the same geometry as in  
Fig. \ref{fig:disorder-operators}). The monopole operator $\tau_3({\tilde {\bm r}})$ anticommutes with the plaquette operator (the second term of the Hamiltonian) which we  denote by
\begin{equation}
\tau_1({\tilde {\bm r}})=\sigma^3_1(\bm r) \sigma^3_2({\bm r}+{\bm e}_1)\sigma^3_2(\bm r) \sigma^3_1({\bm r}+{\bm e}_2)
\end{equation}
Thus, the action of the operator $\tau_3({\tilde {\bm r}})$  is to create (and destroy)  a $\pi$ flux excitation 
at the plaquette labeled by the dual site ${\tilde {\bm r}}$, and $\tau_1({\tilde {\bm r}})$, with eigenvalues $\pm 1$, measures the flux. 

On the other hand, the monopole operator has an expectation value in the confining phase, 
and has exponentially decaying correlations in the deconfined 
phase where a $\pi$ flux excitation has a finite energy gap.
It is easy to see that these gauge-invariant operators define the duality transformation to the 2+1 dimensional quantum Ising model, with Hamiltonian
\begin{equation}
H=-\sum_{{\tilde {\bm r}};j=1,2} \tau_3({\tilde {\bm r}}) \tau_3({\tilde {\bm r}}+{\bm e}_j)-\lambda \sum_{{\tilde {\bm r}}} \tau_1({\tilde {\bm r}})
\label{eq:H-TFIM}
\end{equation}
Here we used the Gauss law of Eq.\eqref{eq:gauss}. Hence, this duality transformation  is a map from the gauge-invariant sector of the gauge theory onto the quantum Ising model.

Following the same line of logic that leads to the representation of the 2D Ising model and its 1D quantum version in terms of Majorana fermions, discussed in the preceding sections,  it has been possible to construct a fermionic version of the 3D Ising gauge theory and the dual 3D Ising model. Here too, the fermions arise as composite operators of a disorder operator and a spin operator. However, the non-local nature of the disorder operator turns the resulting theory into a lattice theory of  a fermionic string \cite{Fradkin-1979c,Polyakov-1981,Itzykson-1982,Casher-1985,Dotsenko-1987,Orland-1987} which has not yet been understood (and much less solved).

\subsection{Deconfinement and the $\mathbb{Z}_2$ topological phase}
\label{sec:Z2-topo}

As far as the properties of local operators are concerned, although the two descriptions of Eq.\eqref{eq:H-IGT} and Eq.\eqref{eq:H-TFIM} are equivalent, 
they differ in their global properties. Indeed, the two-fold degenerate broken symmetry ground state of the quantum Ising model maps onto the confining phase of the 
Ising gauge theory, which has a unique ground state. Of course, this also happens in the case of the self-dual 1D quantum Ising model. 

More subtle is what happens in the case of the deconfined phase at $\lambda > \lambda_c$ of the 2D $\mathbb{Z}_2$ gauge theory. 
In that case duality maps the unbroken symmetry state of the 2D quantum Ising model to the deconfined phase of the $\mathbb{Z}_2$ gauge theory. 
However,  the deconfined phase of the $\mathbb{Z}_2$ gauge theory  does not break any symmetries and 
yet it has a ground state  degeneracy of topological origin that depends on the genus of the 2D surface:
on a two dimensional surface of genus $g$ the ground state degeneracy of the deconfined phase is four fold degenerate and it is $4^g$. 
This is in fact the simplest example of a topological phase.

The topological degeneracy can be seen most easily in terms of the algebra of the 1-cycles of the (``electric'') Wilson loops and of the (``magnetic'') `t Hooft loops \cite{thooft-1979} on non-contractible spacial loops. Let $W[\gamma_j]$ (with $j=1,2$) be the Wilson loops along the non-contractible 1-cycles along the $j=1$ and $j=2$ directions of the square lattice,
\begin{equation}
W[\gamma_j]=\prod_{\ell \in \gamma_j} \sigma^3(\ell)
\label{eq:Wilson-j}
\end{equation}
which can be regarded as an ``electric'' charge transported around the torus along the non-contractible cycle $\gamma_j$.

 Likewise let ${\tilde W}[\Gamma_j]$ be the `t Hooft magnetic loop on the 1-cycles $\Gamma_j$ of the dual lattice,
 \begin{equation}
 {\tilde W}[\Gamma_j]=\prod_{\ell \in \Gamma_j}\sigma^1(\ell)
 \label{eq:thooft-j}
 \end{equation}
 where the links $\ell \in \Gamma_j$ are pierced by the non-contractible 1-cycle  $\Gamma_j$ of the dual lattice. Similarly, the`t Hooft ``magnetic'' loop can be regarded as a ``magnetic'' charge (a $\pi$ flux) transported around torus along the non-contractible cycle $\Gamma_j$.
 
 It is straightforward to show that these loops satisfy the algebra \cite{Kitaev-2003,Freedman-2004,Hastings-2005,Fradkin-2013}
 \begin{align}
 [W[\gamma_j],W[\gamma_k]]=&[{\tilde W}[\Gamma_j],{\tilde W}[\Gamma_k]]=0\nonumber\\
\{W[\gamma_1],{\tilde W}[\Gamma_2]\}=&\{W[\gamma_2],{\tilde W}[\Gamma_1]\}=0\nonumber\\
\label{eq:W-algebra}
\end{align}
and $W[\gamma_j]^2={\tilde W}[\Gamma_k]^2=1$. Both operators are gauge-invariant and commute with the generator of local $\mathbb{Z}_2$ gauge transformations, defined in Eq.\eqref{eq:gauss}.

Deep in the deconfined phase, $\lambda \to \infty$, where the first term of the Hamiltonian of Eq.\eqref{eq:H-IGT} (the ``string tension term'') is negligible. In this limit the energy of all local excitations is sent to infinity. In this limit, the operators of Eq.\eqref{eq:Wilson-j} and Eq.\eqref{eq:thooft-j}  commute with the Hamiltonian, and either (but not both) can be used to label the states. In this ultra-deconfined limit this system is essentially the same as Kitaev's  ``Toric Code'' \cite{Kitaev-2003}. From the algebra of Eq.\eqref{eq:W-algebra} we deduce that there are four inequivalent states on the 2-torus, and $4^g$ states on a surface of genus $g$. These states are labeled by the eigenvalues of either the Wilson or the `t Hooft loops on non-trivial cycles of the surface. This is the finite-dimensional manifold of topological  states of this phase. 
These properties hold not only in the $\lambda \to \infty$ limit but also  throughout the entire deconfined phase, $\lambda > \lambda_c$, provided that the excitation energy gap remains finite (i.e. inside the radius of convergence of an expansion in powers of $1/\lambda$). 

This exact degeneracy of topological origin is a property of  the thermodynamic limit. In a finite system, the manifold of topological ground states develops an  exponentially small energy gap in system size. Thus, in the thermodynamic limit  duality  is a many-to-one mapping, in this case from the topological manifold of the deconfined gauge theory to the disordered phase of the quantum Ising model. In a finite system duality is, instead,  a mapping between the equal-amplitude superposition of states in the different topological sectors to the states of  the quantum Ising model.

The finite energy excitations of the deconfined phase (with infinite energy gap in the $\lambda \to \infty$ limit) are created by the monopole operator, $\tau_3({\tilde {\bm r}})$, that creates (and destroys) a magnetic charge of flux $\pi$. A state with a $\mathbb{Z}_2$ ``electric'' charge  is instead created (and destroyed) by an open Wilson line along some path $\gamma(\bm r)$ of the lattice, ending at a lattice site $\bm r$:
\begin{equation}
W[\gamma({\bm r})]=\prod_{({\bm r}',{\bm r}'+{\bm e}_k) \in \gamma(\bm r)} \sigma_k^3({\bm r}')
\label{eq:Z2-charge}
\end{equation}
provided that the constraint of Eq.\eqref{eq:gauss} is now equal to $-1$ where the charge operator is inserted. In the model with the Hamiltonian of Eq.\eqref{eq:H-IGT} the energy of a $\mathbb{Z}_2$ ``electric charge has infinite energy. In addition, a composite operator made of a $\mathbb{Z}_2$ electric charge and a $\mathbb{Z}_2$ magnetic charge creates a (Majorana) fermion, much as in the Kadanoff-Ceva construction in the 2D classical Ising model, or in the fermionized version of the Ising gauge theory of Ref. \cite{Fradkin-1979c}.

States carrying the $\mathbb{Z}_2$ charge with finite energy can only be allowed if the theory now includes a dynamical  $\mathbb{Z}_2$ (Ising) matter field on the sites of the lattice. The phase diagram of a theory of this type (and its generalizations to other abelian and non-abelian gauge groups)  was studied by Shenker and me \cite{Fradkin-1979}. We showed that this more general theory has two phases: a phase  connecting smoothly the confinement and the Higgs regimes, and a ``free charge'' phase smoothly connected to the deconfined phase of the gauge theory. There we showed that in the free charge phase the operators that create the states are necessarily non-local. This is nowadays regarded as a tell-tale feature of a topological phase. In retrospect, this was the first,  and simplest, example of a topological state of matter, but this became clear only much later. An interesting feature of the the 2+1 dimensional $\mathbb{Z}_2$ Ising gauge theory with $\mathbb{Z}_2$ matter fields is that it is manifestly self-dual, resulting in a symmetry of the phase diagram. More importantly, under duality electric and magnetic charges are are mapped into each other, while the Hamiltonian retains its form.

Thus, deep in the deconfined phase, the Hilbert space reduces to a finite dimensional space of topological origin. From this perspective, the four degenerate states on a 2-torus are labeled by four ``anyons'': the identity $I$, the ``electric'' charge $e$, the ``magnetic' charge $m$, and the fermion $\psi$ (a composite operator of the electric and the magnetic charges). 

\subsection{Effective Topological Field Theory Description}

We will take a different look at the topological nature of the deconfined phase by using an effective topological field theory, Chern-Simons gauge theory \cite{Witten-1989}, which has been used with great success to describe and explain the topological nature of two-dimensional fractional quantum Hall fluids \cite{Laughlin-1983,Zhang-1989,Lopez-1991,Wen-1990,Wen-1992,Wen-1995}.  

The topological aspects of the deconfined phase can alternatively be described by a  multi-component (abelian) Chern-Simons gauge theory of  an $N$-component gauge field $\mathcal{A}_\mu^I(x)$, with $I=1,\ldots,N$ and Lorentz index $\mu=1,2,3$. The action  is 
\begin{equation}
\mathcal{L}=     \int_{\mathcal M}d^3x \;  \frac{1}{4\pi} K^{IJ}\epsilon^{\mu \nu \lambda} \mathcal{A}^I_\mu(x) \partial_\nu \mathcal{A}^J_\lambda(x)
\label{eq:CS}
\end{equation}
 Here $\epsilon^{\mu \nu \lambda}$ is the Levi-Civita tensor and $K^{IJ}$ is a non-singular, invertible, integer-valued $N \times N$ matrix.  This theory is locally gauge-invariant under  the gauge group  $U(1)^N$.  Invariance under large gauge transformations on non-trivial closed manifolds  holds is  $K^{IJ}$ is an integer-valued matrix \cite{Witten-1989}. The manifold is $\mathcal{M}=\Sigma \times \mathbb{R}$, where $\Sigma$ is a spatial manifold (a disk, a sphere, a torus, etc) and $\mathbb{R}$ is time.

This theory is topological in the sense that the, at the classical level, the action does not depend on the metric of the manifold. Thus, its energy-momentum tensor vanishes identically and, in particular, the Hamiltonian is zero. In this theory the Gauss law  is a local constraint between a  ``charge'' density $J_0^I(x)$ and the  gauge flux $\mathcal{F}^I(x)=\epsilon_{ij} \partial_i \mathcal{A}_j^I(x)$,
\begin{equation}
J_0^I(x)=\frac{1}{2\pi} K^{IJ} \mathcal{F}^J(x)
\label{eq:CS-gauss}
\end{equation}
Thus, the constraint implies that the allowed states are charge-flux composites, particles with fractional statistics known as anyons \cite{Wilczek-1982}. 

For a generic integer-valued matrix $K^{IJ}$, the action of Eq.\eqref{eq:CS}, which is first order in time and space derivatives, is odd under time-reversal and parity (which in two space dimensions is a mirror symmetry). For this reason, in general the allowed statistics are phase factors, e.g. in the case of the Laughlin fractional quantum Hall states of a system of fermions the matrix $K^{IJ}$ is just $1 \times 1$ and it is just  an odd integer $m$, and the statistical phase is $\exp(i\pi/m)$. At the quantum level, the action of Eq.\eqref{eq:CS}  implies that this theory
on a surface $\Sigma$ with non-trivial topology of genus $g$ has a finite-dimensional Hilbert space of dimension $|\textrm{det} K|^g$ (for details, see Ref. \cite{Fradkin-2013}).

The topological properties of the deconfined $\mathbb{Z}_2$ gauge theory in $2+1$ space-time dimensions, discussed in Section \ref{sec:Z2-topo}, have a simple representation in terms of a topological field theory of the form of Eq.\eqref{eq:CS}. Since the $\mathbb{Z}_2$ gauge theory does not  break time reversal and parity, it is described by a Chern-Simons theory  with $N=2$ components,  with a $2 \times 2$ $K$ matrix \cite{Freedman-2004}
\begin{equation}
K^{IJ}=
\begin{pmatrix}
0 & 2\\
2 & 0
\end{pmatrix}
\label{eq:K-Z2}
\end{equation}
This  system has a 4-fold degeneracy on a 2-torus, and  four anyons with the  quantum numbers  of the $\mathbb{Z}_2$ gauge theory.

\section{Particle-vortex duality and its generalizations: fermions and bosons}
\label{sec:generalized}

\subsection{Particle-Vortex Duality and 3D $XY$ Models}

The disordered high temperature phase of the classical spin models that we discussed has a representation as a theory of closed loops (i.e. the high temperature expansion). Regarded as a quantum field theory in imaginary time, the closed loops represent the worldlines of virtual particles of a vacuum (ground) state. In the case of the $XY$ model, which has a complex order parameter, the loops are oriented and represent the worldlines of massive charged particles. On the other hand, in $D=3$ Euclidean dimensions, the low-temperature phase can instead be viewed as a theory of closed {\em vortex loops}. Thus, theories of this type have a {\em particle-vortex duality} which maps the vortices of the broken symmetry phase to the particles of the unbroken phase \cite{Peskin-1978,Thomas-Stone-1978,Dasgupta-Halperin-1981}. Closed loops can be regarded as a set of integer-valued locally-conserved currents $\ell_\mu(x)$. The partition function for the loops has the form, a generalization of the Coulomb gas of the 2D case,
\begin{equation}
Z_{\rm loop}=\sum_{[\ell_\mu]} \exp(-S[\ell_\mu]) \; \delta(\Delta_\mu \ell_\mu(x))
\label{eq:Z-loop}
\end{equation}
The Euclidean action $S[\ell_\mu]$ has the form 
\begin{equation}
S[\ell_\mu]=\frac{1}{2} \sum_{x,y} \ell_\mu(x) G_{\mu \nu}(x-y) \ell_\nu(y)
\label{eq:S-ell}
\end{equation}
In this case there is no self-duality since the vortex loops of the broken symmetry phase have  interactions between the currents and  $G_{\mu \nu}(x-y)$ has a Biot-Savart form at long distances (divided by the dimensionless temperature $T$), while  the particle loops of the high temperature phase have  local interactions and $G_{\mu \nu}(x-y) \simeq  \frac{T}{2} \delta(x-y) \delta_{\mu \nu}$.

Technically, the duality transformation involves solving the constraint and then using the Poisson summation formula to map the problem to the dual loops, a generalization of the  procedure used in 2D by Jos\'e and coworkers \cite{jose-1977}. A review on the general form of this duality is found in Ref. \cite{Savit-1980}. 

In the case of the thermal transition of a superconductor with a thermally fluctuating electromagnetic field, the interactions between the vortices are screened at long distances (due to the Meissner effect), and the theory of loops has a similar form in both phases \cite{Peskin-1978,Dasgupta-Halperin-1981}. This particle-vortex duality has also described successfully the quantum superconductor-insulator transition in disordered systems \cite{Fisher-1990} where it led to the proposal of a universal conductivity at the quantum critical point.

\subsection{Duality in fractional quantum Hall fluids}

A similar loop representation, and the associated particle-vortex duality, has also been used to describe the sequence of fractional and integer quantum Hall plateaus and their transitions \cite{Lee-1989,Lee-Kivelson-Zhang-1991,Kivelson-Lee-Zhang-1992,Lutken-Ross-1992}. In a fractional quantum Hall fluid the low-lying excitations are the Laughlin quasiholes which are fractionally charged vortices of the incompressible fluid and are anyons. On the other hand in the Hall insulator the excitations are electrons. In fact, a remarkable experiment showed that the current-voltage curves of this system near this quantum phase transition have a remarkable symmetry exchanging current with voltage \cite{Shimshoni-Sondhi-Shahar-1997}. This symmetry is natural if the two phases are mapped into each other by particle-vortex duality, which here means exchanging    electric and magnetic charges.

A physical way to understand the duality on quantum Hall fluids is in terms of a hydrodynamic picture \cite{Frohlich-1991,Wen-1995}. For simplicity we will consider only the Laughlin states, with filling fraction $\nu=1/m$ of the lowest Landau level.  A fractional quantum Hall fluid is an incompressible state of electrons in a large magnetic field in two space dimensions. It has a conserved current $j_\mu(x)$ which obeys a continuity equation, $\partial_\mu j^\mu=0$. From the local conservation law it follows that the current can be written as 
\begin{equation}
j_\mu=\frac{1}{2\pi} \epsilon_{\mu \nu \lambda} \partial^\nu a^\lambda
\label{eq:j-dual}
\end{equation}
The hydrodynamic field $a^\mu$ is a gauge field in the sense that a gauge transformation $a_\mu \to a_\mu + \partial_\mu \Phi$ does not change the distribution of currents. In particular, this relation means that the electronic charge density, $j_0$, maps onto the flux of the gauge field $a_\mu$ (up to the factor of $1/(2\pi)$).  Hence, the hydrodynamic theory is the dual of the theory of electrons in the sense of electromagnetic duality. Since the FQH state is incompressible and has a finite energy gap, the low energy effective action is a local functional of $a_\mu$, it must be gauge-invariant and odd under time reversal (due to the magnetic field). These requirements imply that the effective low-energy Lagrangian for $a_\mu$ in the FQH state has the form \cite{Wen-1995}
\begin{equation}
\mathcal{L}[a_\mu]=\frac{m}{4\pi} \epsilon_{\mu \nu \lambda} a^\mu \partial^\nu a^\lambda-\frac{e}{2\pi} A^\mu \epsilon_{\mu \nu \lambda} \partial^\nu a^\lambda+j_v^\mu a_\mu +\ldots
\label{eq:Seff-FQH}
\end{equation}
 The first term is the Chern-Simons term of Eq.\eqref{eq:CS} and its coefficient, $m$ (the ``level'' of the Chern-Simons theory), must be an integer for the theory to be invariant under large gauge transformations \cite{Witten-1989}. The second term is the coupling of the charge current, $-e j_\mu$, to an external electromagnetic field $A_\mu$. The third term represents the coupling of the hydrodynamic field to the worldlines of the vortices of the fluid, represented by the vortex currents $j^\mu_v$. This effective hydrodynamic theory can be regarded as the ``dual'' of the theory of electrons in fractionally filled Landau levels. We will see below that similar effective topological field theories are duals of theories of topological insulators (and topological superconductors) in three space dimensions.

\subsection{Self-Dual Loop Models and Modular Invariance}

Motivated by the apparent self-duality observed in the quantum Hall plateau transitions, Kivelson and Fradkin in 1996 introduced a generalized loop model in which self-duality is a manifest property of the partition function \cite{Fradkin-1996}. The generalized loop model embodies the notion of flux attachment by regarding the loop currents $\ell_\mu$ are the worldlines particles that carry electric charge on the links of a cubic lattice and magnetic charge (flux) on the links to the dual lattice. Thus, the particles of this theory are charge-flux composites, anyons. 

The partition function of this loop model has the  form as Eq.\eqref{eq:Z-loop}. The action $S[\ell]$ has the  form of Eq.\eqref{eq:S-ell} but with the important difference that the kernel  is  complex, $G_{\mu \nu}(x-y)=G_{\mu \nu}^{e}(x-y)+i G_{\mu \nu}^{o}(x-y)$. 
The real part, $G{\mu \nu}^e(x-y)$, that describes the interactions,  is even under parity and time-reversal. The contribution of the imaginary part, $G_{\mu \nu}^o(x-y)$, to the loop Euclidean action is proportional to the linking number of  the loops (the Gauss invariant) which is a topological invariant of the loop configurations. This term describes the fractional statistics of the particles and, as such, it is odd under parity and time-reversal. Since the loops are closed, this theory is invariant under charge conjugation (particle-hole) symmetry and has ``zero density.'' So, this a theory of particles with fractional statistics with two (and opposite) charges.
 In the self-dual model, these kernels have the long distance form (in momentum space)
\begin{equation}
G_{\mu \nu}^{e}(k)=2\pi \frac{g}{\sqrt{k^2}} \left(\delta_{\mu\nu}-\frac{k_\mu k_\nu}{k^2}\right), \qquad
G_{\mu \nu}^{o}(k)=2\pi f\;  \frac{\epsilon_{\mu \nu \lambda}  k_\lambda}{k^2}
\label{eq:kernels}
\end{equation}
where $g$ is a dimensionless coupling constant and $2\pi f$ is the statistical angle of the anyons. By construction, since the linking number of a loop configuration is an integer, the weight of a loop configuration is  periodic in the statistical angle and hence is invariant under $f \to f + n$, where $n \in \mathbb{Z}$. 

It is easy to show that the dual of this loop model is another loop model of the same form but with the dual coupling constants $g_D$ and $f_D$,
\begin{equation}
g_D=\frac{g}{g^2+f^2}, \qquad f_D= -\frac{f}{g^2+f^2}
\label{eq:dual-couplings}
\end{equation}
It is convenient to define the complex variable $z=f+ig$, in terms of which duality is the mapping $S$
\begin{equation}
S: z \to z_D=-\frac{1}{z}
\label{eq:complex-duality}
\end{equation}
The invariance under periodic shifts of the statistical angle is the mapping $T$
\begin{equation}
T: z \to z+1
\end{equation}
The mappings $S$ and $T$ do not and generate  the infinite discrete non-abelian group of fractional linear  transformations $SL(2,\mathbb{Z})$
\begin{equation}
\mathcal{T}:z=\frac{az+b}{cz+d}
\label{eq:SL2Z}
\end{equation}
where $a, b, c, d \in \mathbb{Z}$ and $ad-bc=1$.
In Ref. \cite{Fradkin-1996} it was shown that the loop partition function is invariant under these modular transformations (see also Ref.\cite{Burgess-Dolan-2001} for applications of the modular group to quantum Hall systems). 

The coupling of the loop model to an external background electromagnetic field $A_\mu$ breaks self-duality explicitly. However, the complex quantity $D(z)$, that parametrizes the current correlation functions, has a simple transformation law under $SL(2,\mathbb{Z})$ \cite{Fradkin-1996}:
\begin{equation}
D(-\frac{1}{z})=z^2D(z)+z, \qquad D(z+1)=D(z)
\label{eq:D}
\end{equation}
Since under charge conjugation $(f,g) \to (-f,g)$, time-reversal symmetry requires that $D(z)$ obeys the reflection symmetry $D(-z^*)=-D^*(z)$.
The quantity $D(z)$ that obeys the (inhomogeneous) transformation law of Eq.\eqref{eq:D} is known as the anomalous modular form of weight $2$.  It follows that if $z_0$ is a fixed point under $SL(2,\mathbb{Z})$ (or, more precisely, under a subgroup), then $D(z_0)=i/(2 \textrm{Im} z_0)$. As a result, these fixed points  define self-dual systems with {\it finite, universal,  conductivity} $\sigma_{xx} \neq 0$, and vanishing Hall conductivity, $\sigma_{xy}=0$. 
Systems of this type must be at a quantum critical point. 

Two types of fixed points were identified in Ref. \cite{Fradkin-1996}: a) {\em bosonic} fixed points, with $z_0=i$ (and its periodic images), b) {\em fermionic} fixed points along lines $z=\frac{1}{2}+ig$. For example, the longitudinal conductivity at the bosonic fixed point $z_0=i$ was found to be $\sigma_{xx}=\frac{1}{2} \frac{e^2}{2\pi}$, the  value conjectured at the superconductor-insulator quantum critical point \cite{Fisher-1990}. Another example is the fermionic fixed point $z_0=\frac{1}{2}+i \frac{\sqrt{3}}{2}$ the value of the longitudinal conductance is $\sigma_{xx}=\frac{1}{\sqrt{3}} \frac{e^2}{2\pi}$. In addition to finite fixed points, $SL(2,\mathbb{Z})$ has fixed points at extreme values $g \to \infty$ and $g \to 0$ which describe, respectively, an infinite number of gapped insulating phases in which the loops are suppressed, and (also gapped) phases in which the loops proliferate, with non-trivial statistics, separated by quantum critical points. 

Modular invariance also arises in $U(1)$ abelian gauge theories in 3+1 dimensions with a $\theta$ (axion) term \cite{Witten-1995,Witten-2003}. The (Euclidean) action of this gauge theory is 
\begin{equation}
S[a_\mu]=\int_{\mathcal{M}}d^4x \left(-\frac{1}{4g_M^2} F_{\mu \nu}^2+i \frac{\theta}{32\pi^2} \epsilon_{\mu \nu \lambda \rho}F^{\mu \nu} F^{\lambda \rho}\right)
\label{eq:Maxwell+axion}
\end{equation}
The first term is the Maxwell action of electromagnetism and $g_M^2$ is the coupling constant (i.e. $e^2$). On a closed manifold $\mathcal{M}$ the second term, the $\theta$ term,  is proportional to the instanton number,  and the coupling constant $\theta$ is the ``axion''. However, on an open manifold whose boundary is the two-dimensional closed spatial surface $\Sigma$, the axion term integrates to a Chern-Simons term on $\Sigma \times \mathbb{R}$ (here $\mathbb{R}$ denotes time), with Chern-Simons level $k=\frac{\theta}{2\pi}$. This theory has a modular invariance where the modular parameter is $\tau=\frac{\theta}{2\pi}+i \frac{4\pi}{g_M^2}$ \cite{Witten-1995}. 

Since the self-dual loop model in 2+1 dimensions and the $U(1)$ gauge theory are both modular invariant it is natural to seek a connection between them. Indeed, one can couple the $U(1)$ gauge theory to a bosonic matter field defined on the boundary $\Sigma \times \mathbb{R}$ of $\mathcal{M}$. This a situation that occurs in topological insulators (where the matter field is fermionic, as we will see below). The connection with the loop model is achieved in its first-quantized formulation where the matter field is represented by a sum over closed loops (the worldlines).
The expectation value of a Wilson loop of charge $q$ of this theory on a closed path $\gamma$ in 3+1-dimensional Euclidean space time is
\begin{equation}
\Big\langle \exp \left(iq \oint_\gamma dx_\mu A_\mu \right) \Big\rangle=\exp \left( -\frac{q^2}{2} \oint_\gamma dx_\mu \oint_\gamma dy_\nu G_{\mu \nu}(x-y) \right)
\label{eq:Wilson-3+1}
\end{equation}
where $G_{\mu \nu}(x-y)=\langle A_\mu (x) A_\nu(y)\rangle$ is the (Euclidean) propagator of the gauge field $A_\mu$ in the theory of Eq.\eqref{eq:Maxwell+axion}. When the  loops $\gamma$ are restricted to a 2+1-dimensional manifold  $\Sigma \times \mathbb{R}$, the expectation value of Eq.\eqref{eq:Wilson-3+1} yields  the same as the kernel of the loops in the self-dual theory of Ref.\cite{Fradkin-1996}. 

There is, however, a subtle difference in the way the modular group acts in the two theories. In the loop model, the transformation $T: z\to z+n$ is simply the invariance under periodic shifts of the statistical angle of the anyons. Instead, in the $U(1)$ gauge theory with a $\theta$ term, the transformation $T: \tau \to \tau+n$ is just the  shift of the $\theta$ parameter by $2\pi n$. This is a symmetry in virtue of the quantization of the instanton number of the $3+1$-dimensional theory. However, these two transformations are not physically equivalent since the latter shifts the Chern-Simons level and hence changes the statistics.

\subsection{Fermionic Duality and Bosonization}

In this section we will discuss, albeit briefly and qualitatively, the relation between the ideas presented in previous sections and current work on duality in relativistic fermionic and bosonic theories, which have important implications for systems in condensed matter physics. 
Originally formulated for 1+1-dimensional systems, their extension  to higher dimensions are interesting and have a long history. Recent impetus for developing these correspondences in general dimension has come from the discovery of materials known as topological insulators whose electronic structure is well described, at low energies, by Dirac and Weyl fermions \cite{Hasan-Kane-2010,Qi-Zhang-2011}.
A  clear and comprehensive presentation of the current work on this problem can be found in Ref. \cite{Seiberg-Senthil-Wang-Witten-2016}.

\subsubsection{Fermion-Boson Duality in 1+1 Dimensions}

The earliest version of fermion-boson duality is the well known bosonization of fermionic systems in 1+1 space-time dimensions \cite{Lieb-1965,Luther-Emery-1974,Coleman-1975,Mandelstam-1975} (for a pedagogical presentation see Ref. \cite{Fradkin-2013}). This mapping is a fundamental tool for the understanding  interacting fermionic systems in 1+1 dimensions, quantum spin chains, and QED$_2$ and QCD$_2$, as well as the edge states of quantum Hall systems. Bosonization is a mapping, at the operator level, between a system of free massless Dirac fermions  $\psi_a(x)$, a two-component spinor field with Lagrangian $\mathcal{L}_F$, and a compactified massless boson $\phi(x)$, with Lagrangian $\mathcal{L}_B$,
\begin{align}
\mathcal{L}_F={\bar \psi}(x) i(\slashed{\partial}+i\slashed{A}) \psi(x) &\leftrightarrow \mathcal{L}_B=\frac{1}{8\pi} \left(\partial_\mu \phi(x)\right)^2+\frac{1}{2\pi}  \epsilon_{\mu\nu}  \partial^\mu \phi A^\nu
\label{eq:duality-1D-Lagrangians}\\
j_\mu={\bar \psi(x)} \gamma_\mu \psi(x) & \leftrightarrow \frac{1}{2\pi} \epsilon_{\mu \nu}\partial^\nu \phi(x)
\label{eq:duality-1D-currents}
\end{align}
where $A^\mu$ is an external gauge field. The boson is compactified as $\phi(x+L)=\phi(x)=2\pi N$, with $N\in \mathbb{Z}$ and $L$ is the length. In this correspondence, the fermion current $j_\mu$ is mapped onto the ``topological current'' of the boson, Eq.\eqref{eq:duality-1D-currents}, which is characteristic of a duality. 
The compactification of the boson $\phi$ is a consequence of charge quantization: the total fermion number is $N_F=\int_0^L dx  j_0(x)=\frac{1}{2\pi} \int_0^L dx \partial_x\phi=\frac{\Delta \phi}{2\pi}=N $. Thus, the winding number $N$ of the boson labels a sector of the fermionic theory with fermion number $N_F=N$. Furthermore, fermion operators are mapped onto  soliton (vertex) operators in the bosonic theory \cite{Mandelstam-1975}. These soliton (or kink) operators are the continuum version of the  kink (disorder) operators of earlier sections. It is important to stress that the fermion-boson equivalency  holds not only at the level of operators but also at the level of the spectrum, and of the partition functions. 

\subsubsection{Fermion-Boson Duality in 2+1 Dimensions}

In the case of massive Dirac fermions of mass $M$,  in 2+1 dimensions the fermion-boson mapping takes the form of an effective low-energy, hydrodynamic, bosonic Lagrangian of the form \cite{Fradkin-Schaposnik-1994,Burgess-Quevedo-1993,LeGuillou-1997,Chan-Hughes-Ryu-Fradkin-2013}
\begin{equation}
\mathcal{L}_{\rm eff}[a_\mu, b_\mu]=\frac{1}{2\pi} \epsilon_{\mu \nu \lambda} A^\mu \partial^\nu b^\lambda -\frac{1}{2\pi} \epsilon_{\mu \nu \lambda} a^\mu \partial^\nu b^\lambda+\frac{K}{4\pi} \epsilon_{\mu \nu \lambda} a^\mu \partial^\nu a^\lambda +\ldots
\label{eq:bosonization-2+1-massive}
\end{equation}
where we have neglected a subdominant, non-topological, Maxwell-type term with a dimensionful prefactor $\propto 1/M$. Here $A_\mu$ is a background electromagnetic field.
For a theory with $N_f$ types (``flavors'') of two-component Dirac spinors, the  coefficient  of the Chern-Simons term (the third term of this Lagrangian)  is $K=\textrm{sign}(M) N_f/2$. In particular, for topological Chern insulators $K$ is equal to the Chern number of the fully occupied states. On the other hand, the case of an odd number of Dirac fermions, $N_f$ odd, is subtle, since in this case the Chern-Simons level is a half-integer which violates the requirement of invariance under local and large gauge transformations on a closed manifold \cite{Witten-1989}. We will return to this question below. The second term of Eq.\eqref{eq:bosonization-2+1-massive} is known as the $BF$ term and it is also topological.

From the Lagrangian of Eq.\eqref{eq:bosonization-2+1-massive} it follows that the fermion current, $j_\mu={\bar \psi} \gamma_\mu \psi$, has the bosonized form
\begin{equation}
j_\mu \leftrightarrow \frac{1}{2\pi} \epsilon_{\mu \nu \lambda} \partial^\nu b^\lambda
\label{eq:bosonized-current-2+1}
\end{equation}
This identification embodies the electric-magnetic nature of the duality: charge $\leftrightarrow$ flux. This equation also implies that a fermion operator acting at some space-time point $x$ is equivalent to having a magnetic monopole (which in 2+1 dimensions is an instanton) at $x$ of the gauge field $b_\mu$.

One may ask, do these identifications apply to the case of massless fermions? This turns out to be a subtle problem and much of what is known about it is at the level of conjectures which have survived many non-trivial checks (see, e.g. Ref. \cite{Seiberg-Senthil-Wang-Witten-2016}). For the sale of conciseness, here  we have purposely ignored many technically important details, particularly on the key role of boundary conditions and of the types of manifolds.  For more details (and generalizations) of this mapping see Ref.\cite{Seiberg-Senthil-Wang-Witten-2016} (and references therein) whose treatment we followed closely.

Recent work  on this problem has led to may conjectured dualities \cite{Seiberg-Senthil-Wang-Witten-2016,Karch-Tong-2016,Metlitski-Vishwanath-2016} (extending earlier work of Refs. \cite{Aharony-Gur_Ari-Yacoby-2012,Jain-Minwalla-Yokoyama-2013} in some large-$N$ limits). Seiberg and coworkers \cite{Seiberg-Senthil-Wang-Witten-2016} recast the duality results by Peskin \cite{Peskin-1978} and by Dasgupta and Halperin \cite{Dasgupta-Halperin-1981} as a mapping between the Wilson-Fisher fixed point of a complex scalar field $\phi$ (minimally coupled to a background gauge field $A_\mu$) to a dual complex scalar field $\widehat \phi$ coupled to a dynamical gauge field $a_\mu$. Schematically, the correspondence between the Lagrangians is
\begin{equation}
|(\partial_\mu+iA_\mu)\phi |^2-|\phi |^4 \leftrightarrow |(\partial_\mu+ia_\mu){\widehat \phi})|^2-|{\widehat \phi}|^4+\frac{1}{2\pi} \epsilon_{\mu \nu \lambda} A^\mu \partial^\nu a^\lambda
\label{eq:particle-vortex-scalar}
\end{equation}
where the (renormalized) coefficient $r$ of the $\phi^2$ term and $\widehat r$ of the $|\widehat \phi |^2$ terms have been tuned to zero (the critical point) and the coefficients of $|\phi |^4$ and $|\widehat \phi |^4$ have been tuned to the Wilson-Fisher fixed points. Particle-vortex (or electric-magnetic) duality is manifest in the form of the coupling to the electromagnetic field $A_\mu$ in the mapping of Eq.\eqref{eq:particle-vortex-scalar}, and that the conserved current of the bosons has the same duality expression as in Eq.\eqref{eq:bosonized-current-2+1} (in terms of the gauge field $a_\mu$).

The duality for massless Dirac fermions in 2+1 dimensions is more subtle. Largely inspired by the results of Refs. \cite{Aharony-Gur_Ari-Yacoby-2012,Jain-Minwalla-Yokoyama-2013}, Seiberg and coworkers \cite{Seiberg-Senthil-Wang-Witten-2016} conjectured that a similar duality applies for a theory with a single Dirac fermion with a global $U(1)$ symmetry to a boson (charged scalar field) at its Wilson-Fisher fixed point coupled to a $U(1)$ (dynamical) Chern-Simons gauge field $b_\mu$:
\begin{equation}
i{\bar \psi} \left( \slashed{\partial}+i\slashed{A} \right) \psi 
\leftrightarrow 
|\left(\partial_\mu+i b_\mu \right)\phi |^2-|\phi |^4
+\frac{1}{4\pi} \epsilon_{\mu \nu \lambda} b^\mu \partial^\nu b^\lambda
+\frac{1}{2\pi} \epsilon_{\mu \nu \lambda} b^\mu \partial^\nu A^\lambda
\label{eq:fermion-boson-duality-2+1}
\end{equation}
The validity of this mapping has been checked by identifying operators on both sides, most often this can be done deep in some phase and then extrapolated back to the massless/critical case. Contrary to the 1+1-dimensional case, the identification of the partition functions is still lacking, given the the bosonic theory is at a non-trivial fixed point.
Once again, electromagnetic duality is apparent in this mapping from the form of the coupling to the background electromagnetic in the bosonized theory in Eq.\eqref{eq:fermion-boson-duality-2+1} which has the same identification as in  the massive case, c.f. Eq.\eqref{eq:bosonized-current-2+1}.

There is, however, an important subtlety in this case. A free massless Dirac fermion has a parity anomaly: parity (and time-reversal invariance) is violated in any gauge-invariant definition of this theory. A correct definition of this theory (which restores the time-reversal invariance of the free massless Dirac fermion) requires that we add to the left-hand-side of Eq.\eqref{eq:fermion-boson-duality-2+1} a half-quantized Chern-Simons term for the background gauge field $A_\mu$ of the form $-\frac{1}{8\pi} \epsilon_{\mu \nu \lambda} A^\mu \partial^\nu A^\lambda$, which restores time-reversal invariance. (On a generic manifold, a gravitational Chern-Simons term must also be included). The microscopic origin of this term depends of the definition of the theory. In a lattice model (of the type used in topological insulators) it arises form the contribution of the massive fermion ``doublers'' (or, equivalently, from massive Pauli-Villars regulators). Alternatively, as will see below, this theory can be regarded as the boundary of a 3+1-dimensional system.

Another check on the duality follows from adding a mass term, $m_F {\bar \psi} \psi$, to the Dirac theory, which break time-reversal invariance explicitly. Under duality this term maps onto a mass term for the charged boson  {\em of opposite sign} $-m_B^2 \phi^2$. It follows that for $m_F>0$, the massive Dirac theory has a vanishing Hall conductivity, $\sigma_{xy}=0$ (and it is a trivial insulator), whereas for the opposite sign, $m_F <0$, it has a quantized Hall conductivity $\sigma_{xy}=1$ (in units of $e^2/h$). On the bosonic side, the first case maps onto the Higgs phase of the boson, in which the Chern-Simons term of the gauge field $b_\mu$ is inoperative. In contrast, the second case maps onto the symmetric phase of the charged boson in which the $U(1)$ symmetry is unbroken and the boson is massive: this is a topological phase described by a Chern-Simons theory $U(1)_1$. Finally, the Dirac fermion is mapped onto a composite operator made of the boson and a monopole of the gauge field $b_\mu$ with unit magnetic charge. This operator can be regarded as   a disorder operator of the bosonic theory.

We close this discussion by noting that while the Dirac fermion is free, the bosonic theory is at a non-trivial fixed point. In the absence of the Chern-Simons gauge field $b_\mu$, the Wilson-Fisher fixed point is strongly interacting and, more importantly, it has operators, such as the field $\phi$ itself,  with a finite (albeit small) anomalous dimension. Whether this gauged version this theory with a Chern-Simons term also has gauge-invariant operators with non-trivial anomalous dimensions is presently not understood, although there is some evidence in the large-$N$ version of this theory. 

\subsubsection{Fermion-Boson Duality in 3+1 Dimensions}

We close the discussion of fermion-boson duality with a brief discussion of this mapping in 3+1 dimensions. For brevity we will discuss only the case of massive fermions. 
Although this problem has a long history in high-energy physics, much of the current interest originated from the discovery of topological insulators in three space dimensions (see Refs. \cite{Hasan-Kane-2010,Qi-Zhang-2011}.)

In its simplest version this problem is the $\mathbb{Z}_2$ topological insulator, which is represented as a Dirac fermion (a four-component spinor) with a spatially-varying mass which is positive, $m>0$, outside a region $\Omega$ of 3D space  and $m<0$ inside $\Omega$. The Dirac fermion mass vanishes (smoothly) at the boundary $\Sigma=\partial \Omega$ of the region. It is well known that in this case the low-energy states consist is massless of a two-component Dirac spinor at the surface $\Sigma$. From the discuss of the preceding Subsection we gather that this theory has an anomaly at the surface $\Sigma$. If the region $\Omega$ is taken to be an infinitely wide slab of finite (but large) thickness, then there are two massless Dirac  bispinors at each surface with opposite chirality (in the four-dimensional sense). Since the system as a whole is gauge and time reversal invariant, the anomaly of one surface is exactly cancelled by the anomaly of the other surface. This phenomenon is known as the anomaly inflow \cite{Callan-1985}. 

From the perspective of the bulk region $\Omega$, we can find a dual of the massive Dirac fermion (with negative mass $m<0$) in terms of a hydrodynamic effective action, analogous to Eq.\eqref{eq:bosonization-2+1-massive}. Using the same bosonization approach as before, the resulting dual theory involves an anti-symmetric tensor (Kalb-Ramond) gauge field $b_{\mu \nu}$ and a gauge field $a_\mu$ \cite{Chan-Hughes-Ryu-Fradkin-2013}
\begin{equation}
\mathcal{L}=-\frac{1}{2\pi} \epsilon_{\mu \nu \lambda \rho} b^{\mu \nu} \partial^\lambda (a^\rho-A^\rho)+\frac{\theta}{32\pi^2} \epsilon_{\mu \nu \lambda \rho} f^{\mu \nu} f^{\lambda \rho}-\frac{1}{4g_M^2}f_{\mu \nu} f^{\mu \nu}
\label{eq:bosonization-3+1-massive}
\end{equation}
where $A^\mu$ is an external, background, electromagnetic field and $f_{\mu \nu}=\partial_\mu a_\nu-\partial_\nu a_\mu$. Here $g_M$ is an effective (Maxwell) coupling constant and the $\theta$ angle is $\theta=\pi$ for the $\mathbb{Z}_2$ topological insulator and $\theta=0$ for the trivial one. This theory  invariant under $SL(2,\mathbb{Z})$ modular transformations \cite{Witten-1995}.

The duality of Eq.\eqref{eq:bosonization-3+1-massive}  leads to the electro-magnetic duality in terms of the current with the identification
\begin{equation}
j_\mu={\bar \psi} \gamma^\mu \psi \leftrightarrow \frac{1}{2\pi} \epsilon_{\mu \nu \lambda \rho} \partial^\nu b^{\lambda \rho}
\label{eq:current-bosonization-3+1}
\end{equation}
The $\theta$ term is a total derivative and hence it integrates to the boundary where it contributes with a Chern-Simons term of the form $\pm \frac{\theta}{8\pi^2} \epsilon_{\mu\nu \lambda} a_\mu \partial_\nu a_\lambda$ (restricted to the manifold $\Sigma \times \mathbb{R}$), which, for $\theta= \pi$ (where $\pm$ denotes the orientation of $\Sigma$), has precisely the form required by the anomaly.


\begin{acknowledgements}

I am  grateful to Paul Fendley for many enlightening discussions on parafermions and for sharing his unpublished results with me.
This work was supported in part by the National Science Foundation through the grant DMR 1408713 at the University of Illinois.
\end{acknowledgements}


\end{document}